\documentclass[onecolumn,amsmath,amssymb,floatfix,10pt,prd,superscriptaddress,nofootinbib]{revtex4}

\usepackage[parfill]{parskip}    % Activate to begin paragraphs with an empty line rather than an indent
\usepackage{graphicx}
\usepackage{amssymb}
\usepackage{epstopdf}
\usepackage{color}
\usepackage{float}
\usepackage{amsmath}
\usepackage{orcidlink}
\usepackage{latexsym}
\usepackage{amsmath}
\usepackage{amssymb}
\usepackage{amsfonts}
\usepackage{subcaption}
\usepackage[font=small,labelfont=bf,
   justification=justified,
   format=plain]{caption}

\begin{document}

\title{Thermodynamic aspects of FLRW Universe in Einstein-Gauss-Bonnet domains }

\author{Joel F. Saavedra\orcidlink{0000-0002-1430-3008}}
\email{joel.saavedra@pucv.cl}
\affiliation{
Instituto de Física, Pontificia Universidad Católica de Valparaíso, Casilla 4950, Valparaíso, Chile.}

\author{Francisco Tello-Ortiz
 \orcidlink{0000-0002-7104-5746}}
\email{francisco.tello@pucv.cl}
\affiliation{
Instituto de Física, Pontificia Universidad Católica de Valparaíso, Casilla 4950, Valparaíso, Chile.}

\begin{abstract}

For an FLRW model, thermodynamic phase transitions are investigated in the Einstein-Gauss-Bonnet gravity framework. Using the work density, the equation of state is derived, and the criticality conditions are employed to determine the critical points where possible phase transitions occur. The appearance of phase transitions strongly depends on the space-time dimension $n$. In this concern, for $n=5$, there is an ``inverted'' first-order phase transition, where the Gibbs free energy presents a swallow-tail behavior. On the other hand, for $n=6$, the system does not exhibit first order phase transition. In such a case, the Gibbs free energy presents a cusp with stable and unstable branches. 
For the present study, the mentioned phenomena are present for an expanding cosmology, where the matter distribution filling the Universe corresponds to a speculative matter distribution with an equation of state parameter greater than one. 
Interestingly, there are no phase transitions for dimensions greater than $n=6$, nor for expanding or contracting cosmological scenarios. To gain more insights into the system, the microstructure is analyzed using thermodynamic geometry to quantify the normalized scalar curvature. This invariant shows that an attractive  interaction dominates the phase-transition region. Additionally, the topological thermodynamic analysis was performed employing Duan's off-shell map. This study reveals that for $n=5$ we observe a winding number interchange twice, indicating an unstable small/large branch phase transition through an intermediate stable phase. For $n=6$ the number of exotic defects
is one. Consequently, we observe a stable small branch and an unstable large branch.

\end{abstract}
%\date{}

\maketitle

\section{Introduction}

The pioneering works about the thermodynamics description of black holes\footnote{See \citep{Wald:1999vt,Altamirano:2014tva} for a complete revision about black hole thermodynamics.} (BH) in the framework of General-Relativity (GR)
\citep{Bardeen:1973gs,Bekenstein:1973ur,Smarr:1972kt,
Hawking:1975vcx,Gibbons:1976ue,York:1986it}, were the \emph{corner stone} to put forward the so-called gravity-thermodynamic conjecture \citep{Jacobson:1995ab,Padmanabhan:2002sha,Padmanabhan:2003gd}. With these studies, their features have received a boost of new interest, leading to a possible unrevealed and expected unification among GR,
quantum theory and statistical physics. 
Concerning the connection between GR and quantum mechanics, beyond the static scope, the back reaction problem for dynamical BHs was developed in \citep{Kodama:1979vn}, where instead of having Killing vector fields and event horizons, one has a dynamical vector field (later on coined as Kodama's vector field) and apparent horizons. Interestingly, this vector field is associated with a conserved charge, the Misner--Sharp mass \citep{Misner:1964je}. This antecedent led to Hayward to develop the thermodynamic description of dynamical BHs, proposing the so-called thermodynamic unified first law \citep{Hayward:1993wb,Hayward:1994bu,Hayward:1997jp}.

With the thermodynamic description for dynamical systems at hand (which includes the static case), the study of the gravitation--thermodynamics relationship went beyond the study of BHs, expanding to other systems and gravitational models such as cosmological models. Interestingly, the widely accepted cosmological model nowadays, the FLRW model, has been proven to have an apparent horizon \citep{Cai:2005ra,Gong:2007md,Faraoni:2011hf,Faraoni:2015ula}. This main ingredient is the cause of having self-consistent thermodynamics. Furthermore, this feature strongly suggests that the
relationship between the unified first law of thermodynamics of the apparent horizon and the Friedmann
the equation is not just a simple coincidence but rather a deeper  physical connection. Nevertheless, for an FLRW model subject to a perfect fluid matter distribution in the context of GR, no thermodynamics phase transitions were found \citep{Abdusattar:2021wfv}. Therefore, it was necessary to go beyond the scope of GR to obtain some insights about the thermodynamics description of the FLRW Universe \citep{Kong:2021dqd,Abdusattar:2023pck,Abdusattar:2023hlj}. Mainly, the studies provided in \citep{Kong:2021dqd,Abdusattar:2023pck,Abdusattar:2023hlj}, concern a special four--dimensional tensor-scalar gravity theory belonging to the Horndeski class \citep{Fernandes:2021dsb}.

Taking into account the above antecedents, it is legitimate to ask what role high dimensions play in the thermodynamics description of the Universe in an FLRW background. As a starting point to tackle this issue, in this article, we study the thermodynamics of an FLRW cosmological model in the framework of Einstein-Gauss-Bonnet gravity theory. As it is well-known, this theory belongs to the class of Lovelock gravity theories \citep{Lovelock:1971yv}, containing a plethora of good properties such as: i) it is free from ghost fields, ii) well-established spherically symmetric vacuum solution\footnote{In \citep{Cai:2003kt}, a thoroughly study about BHs thermodynamics was performed in the realm of Lovelock's gravity theory. Of course, this analysis contains the Einstein-Gauss-Bonnet case, since this latter is part of the Lovelock's theory. } \citep{Boulware:1985wk,Cai:2001dz}, iii) well--defined entropy \citep{Myers:1988ze,Cvetic:2001bk,Clunan:2004tb}, iv) Misner-Sharp mass \citep{Maeda:2006pm,Maeda:2007uu,Maeda:2008nz,Maeda:2011ii,Nozawa:2007vq} and v) Friedmann equations \citep{Cai:2005ra,Akbar:2006kj}. 
So, employing the results\footnote{It should be pointed that, recently in \citep{Nojiri:2023wzz} was proposed a modified unified first law (a generalization of Hayward's work), where the work density $W$ only is determined in terms of the density of the fluid $\rho$. Besides, in \citep{Nojiri:2022nmu} were proposed more general cosmological field equations, obtained from the thermodynamic of the apparent horizon throughout the proposal of a three-parameter entropy--like quantity: \begin{equation}\nonumber
S_{\mathrm{C}}(\delta, \beta, \gamma)=\frac{1}{\gamma}\left[\left(1+\frac{\delta}{\beta} S_{\mathrm{BH}}\right)^\beta-1\right].
\end{equation}
Nevertheless, as a first step, we are going to use previous results obtained by Hayward, leaving this new proposal for future works. } provided in \citep{Hayward:1993wb,Hayward:1994bu,Hayward:1997jp}, the equation of state for an FLRW metric is derived. The thermodynamics portrayal comes after clarifying the final sign of the surface gravity and consequently of the temperature \citep{Binetruy:2014ela,Helou:2015yqa,Helou:2015zma}. Here, it is found that the FLRW is characterized by an outer--past apparent horizon, leading to an expanding cosmology with negative temperature. Contrary to the belief that only gravitational systems with positive temperatures are allowed, at least from a theoretical point of view, systems with negative temperatures are admissible. They should not necessarily be discarded \citep{Vieira:2016lyj}. Furthermore, this system has a peculiar characteristic: its matter distribution corresponds to the so-called \emph{stiff matter}. Moreover, it has been shown that this type of matter could play a leading role in the Universe's early evolution \citep{Banks:2008ep}.

Interestingly, higher dimensions play a major role in determining possible thermodynamics phase transitions in the present study. In this regard, only five and six space-time dimensions are allowed. For the former, the system undergoes an ``inverted'' first-order phase transition, while for six space-time dimensions, there are no phase transitions. Instead, the system has well-defined stable and unstable branches. To get more insights about the thermodynamics description of the cosmological system, we also perform the study from the point of view of the so-called geometrothermodynamics or Ruppeiner's geometry \citep{Ruppeiner:1981znl,Ruppeiner:1983zz,Ruppeiner:1995zz,Ruppeiner:2013yca}  and critical phenomena \citep{Goldenfeld:1992qy}. The crucial point of geometrothermodynamics is  to extract, phenomenologically
or qualitatively, the microscopic interaction information of a given system based on  axioms of thermodynamics. The foremost idea lies in the scalar curvature of Ruppeiner's metric accounting for the nature of interactions
among the constituent particles. So, for those systems where microstructures interact attractively, the scalar curvature is positive in nature, whereas it is negative for
predominantly repulsive forces. In this case, for five space-time dimensions, there is a region where this invariant is positive, the region where the phase transition is taking place, and negative for the region where no phase transition is occurring. On the other hand, for six space-time dimensions, this parameter is negative for unstable states and positive for stable states. Considering the critical exponents, $\{\hat{\alpha},\hat{\beta},\hat{\gamma},\hat{\delta}\}$, these have the usual values, recognized as universal values\footnote{In the study of BHs thermodynamics within the framework of gravity theories beyond GR, it has shown that these exponents are not taking the universal values at all \citep{Dehghani:2022gwg}.}. However, as this is a not conventional first order phase transition, Maxwell's construction is not possible, then the critical exponent $\hat{\beta}$ cannot be determined. To further validate our analysis, using Duan's procedure \cite{Duan:1984ws} we explore the topological description \cite{Wei:2021vdx,Wei:2022dzw}, finding a generating point for $n=5$ with a topological charge equal to zero and not generating point for $n=6$ with a topological charge equal to one.

The article is organized as follows: In Sect. \ref{sec2} a general review of the main aspect of the Einstein--Gauss--Bonnet are provided. Besides, the equation of state for an FLRW metric is obtained from the thermodynamics unified first law, and a deep discussion about the final sign of the surface gravity and temperature is given, accompanied by an analysis of the causal nature of the apparent horizon and the possible matter distribution filling the Universe, next, in Sect. \ref{sec3}, $P-v$ phase transitions are explored, providing a complete study about the incidence of the space-time dimensions on this phenomenon. Also, it provides a thorough assay of the ``inverted'' first-order phase transition undergone by the system in Sect. \ref{sec4}, the geometrothermodynamics and critical exponents are analyzed. For the former, we perform a detailed study about the sign of the normalized scalar curvature, accounting for phase transitions near the critical point. The critical exponents for the heat capacity at constant volume $C_{v}$, shear viscosity $\eta$, compressibility $\kappa_{T}$ and pressure $P$ are obtained. In Sect. \ref{sec5}, we study the Duan's topological approach for phase transitions and finally, in Sect. \ref{sec6} we conclude the present investigation.

\section{Einstein--Gauss-Bonnet gravity and the equation of state}\label{sec2}

This section is devoted to the presentation of the main features of the Einstein-Gauss-Bonnet gravity theory. Mainly, it presents its field equations on an FLRW background, the associated Misner--Sharp energy, and thermodynamics properties obtained from the thermodynamics unified first law. This latter allows us to get the equation of state, which is important to the study of phase transitions given in Sect. \ref{sec3}.

\subsection{Field equations on a FLRW background}

The full action of the $n$--dimensional Einstein-Gauss-Bonnet (EGB) gravity, coupled to matter fields, is given by

\begin{equation}\label{fullaction}
S_{GB}=\frac{1}{16\pi G_{n}}\int_{\mathcal{M}} d^{n} x \sqrt{-g}\left[\mathcal{R}+\alpha\mathcal{G}_{\text{GB}}\right] + \text{B.T.} + S_{m},
\end{equation}

where $\mathcal{R}$ is the $n$--dimensional Ricci scalar, B.T. stands for boundary terms in order to have a well-posed variational principle, $G_{n}$ the $n$--dimensional gravitational constant, $\alpha$ is a constant with units of length squared and $S_{m}$ is the action of the matter fields. The Gauss-Bonnet (GB) term $\mathcal{G}_{GB}$ reads as

\begin{equation}\label{GBinvariant}
\mathcal{G}_{\text{GB}}=\mathcal{R}^2-4 R^{\mu \nu} R_{\mu \nu}+R^{\mu \nu \sigma \gamma} R_{\mu \nu \sigma \gamma}.
\end{equation}

As it is well-known, in the low energy limit case of superstring theory, the coupling  $\alpha$ is related to the inverse string tension being positive definite \citep{Gross:1986iv,Gross:1986mw}. Furthermore, the EGB action is a natural extension of the Einstein theory, in the sense that no derivatives higher than second order appear in the field equations. For this to happen, the numerical coefficient of the $\mathcal{G}_{GB}$ must be $\{1,-4,1\}$ as shown in Eq. (\ref{GBinvariant}). This gravitational theory naturally arises in the context of Lovelock gravity \citep{Lovelock:1971yv}.

From the above (\ref{fullaction}), variations with respect to the metric tensor lead to the following equation of motion.

\begin{equation}\label{fieldequations}
G^{\mu}_{\nu}+\alpha H^{\mu}_{\nu}=8 \pi G_{n} T^{\mu}_{\nu},
\end{equation}
being
\begin{equation}
    G_{\mu\nu}=R_{\mu\nu}-\frac{1}{2}g_{\mu\nu}\mathcal{R},
\end{equation}
the Einstein tensor and
             
\begin{align} \label{GBcontribution}
    H_{\mu \nu}&=2\left(\mathcal{R} R_{\mu \nu}-2 R_{\mu \lambda} R_\nu^\lambda-2 R^{\gamma \delta} R_{\gamma \mu \delta \nu}+R_\mu^{\alpha \gamma \delta} R_{\alpha \nu \gamma \delta}\right) -\frac{1}{2} g_{\mu \nu} \mathcal{G}_{G B}
\end{align}

the contribution coming from the GB term (\ref{GBinvariant}), the so--called Lanczos tensor. As we know, for $n=4$, the tensor $H_{\mu\nu}$ is zero because the GB term (\ref{GBinvariant}) becomes a topological invariant, not affecting the dynamic of the theory. 
Here, the tensor $T^{\mu}_{\nu}$ represents, as usual, the energy-momentum tensor for matter fields obtained from $S_{m}$. We are interested in studying the FLRW cosmological model 
\begin{equation}\label{FLRW}
d s^2=-d t^2+a^2(t)\left(\frac{d r^2}{1-k r^2}+r^2 d \Omega_{(n-2)}^2\right),
\end{equation}
in the arena of EGB theory, we assume that the energy-momentum-tensor is the one describing an isotropic (perfect) fluid
\begin{equation}\label{EMT}
T_{\mu\nu}=\left(\rho+p\right)X_{\mu}X_{\nu}+pg_{\mu\nu},
\end{equation}
where $\rho$ is the density, $p$ the isotropic pressure and $X^{\mu}X_{\mu}=-1$ the four--velocity. The field Eqs. (\ref{fieldequations}) subject to the line element (\ref{FLRW}) lead to the following Friedmann equations in the EGB scenario
\begin{equation}\label{FGB1}
\left[1+\tilde{\alpha}\left(H^{2}+\frac{k}{a^{2}}\right)\right]\left[H^{2}+\frac{k}{a^{2}}\right]=\frac{16 \pi G_{n}}{(n-1)(n-2)} \rho,
\end{equation}
\begin{equation}\label{FGB2}
\left[1+2 \tilde{\alpha}\left(H^{2}+\frac{k}{a^{2}}\right)\right]\left[\dot{H}-\frac{k}{a^2}\right]=-\frac{8 \pi G_{n}}{n-2}(\rho+p) .
\end{equation}
As expected, the above expressions are modified by the presence of the EGB terms throughout the coupling constant $\tilde{\alpha}\equiv (n-3)(n-4)\alpha$ and also by the dimension $n$ of the space-time \citep{Cai:2002bn}, recovering the usual Friedmann equations for the GR scenario when $n=4$.

%In obtaining the above expressions we have used the Ricci's scalar
%\begin{equation}
%\mathcal{R}=6\left(\frac{\dot{a}^{2}}{a^{2}}+\frac{\ddot{a}}{a}+\frac{k}{a^{2}}\right),
%\end{equation}
%and the GB term 
%\begin{equation}
%\mathcal{G}_{\text{GB}}=24\left(\frac{k}{a^{2}}+H^{2}\right)\left(\dot{H}+H^{2}\right).
%\end{equation}

To exploit the thermodynamic description encoded in the field Eqs. (\ref{FGB1})--(\ref{FGB2}), it is better to re--express the line element (\ref{FLRW}) in another fashion. This is in order to extract relevant information. So, the line element (\ref{FLRW}) can be written as a warped product as follows
\begin{equation}\label{warped}
    ds^{2}=h_{ij}dx^{i} dx^{j}+R^{2}d \Omega_{(n-2)}^2.
\end{equation}
In the above expression 
$h_{ij}=\text{diag}\{-1,a^{2}/\left(1-kr^{2}\right)\}$ is the metric on $t-r$ plane (with $k=0,\pm 1$ is the spatial curvature), while $(n-2)$--dimensional unit sphere $d \Omega_{(n-2)}^2$ remains unaltered. Nevertheless, the function $R$, which is the areal radius, contains information of the $t-r$ chart, and it is defined as $R(t,r)\equiv a(t)r$.

So, the dynamical apparent horizon ($AH$), defined as a marginally trapped surface with vanishing
expansion \citep{Faraoni:2011hf,Faraoni:2015ula}, can be obtained from the following the relation 
\begin{equation}\label{EqAH} h^{ij}\nabla_{i}R\nabla_{j}R=0, \quad i,j=t,r,
\end{equation}
where $h^{ij}=\text{diag}\{-1,\left(1-kr^{2}\right)/a^{2}\}$ is the inverse of the metric $h_{ij}$. Solving (\ref{EqAH}) one obtains the following solution for the $AH$
\begin{equation}\label{AHvalue}
    R_{AH}=\frac{1}{\sqrt{H^{2}+\frac{k}{a^{2}}}},
\end{equation}
with $H=\dot{a}/a$ being the Hubble's parameter. It should be noted that for a flat FLRW, that is, when $k=0$, the expression (\ref{AHvalue}) coincides with the Hubble's radius \citep{Faraoni:2015ula}. From now on, we are going to keep the spatial curvature $k$ arbitrary. Therefore, with the solution (\ref{AHvalue}) at hand, the field Eqs. (\ref{FGB1})--(\ref{FGB2}) can be recast as
\begin{equation}\label{FGB11}
\frac{1}{R^{2}_{AH}}+\frac{\tilde{\alpha}}{R^{4}_{AH}}=\frac{16 \pi G_{n}}{(n-1)(n-2)} \rho,
\end{equation}
\begin{equation}\label{FGB22}
\left(1+2 \frac{\tilde{\alpha}}{R^{2}_{
AH}}\right)\left(\dot{H}-\frac{k}{a^2}\right)=-\frac{8 \pi G_{n}}{n-2}(\rho+p).
\end{equation}

It is worth mentioning that, in the limit $n=4$, thus $\tilde{\alpha}\rightarrow 0$, GR does not present phase transitions for a FLRW model \citep{Kong:2021dqd}. Therefore, thermodynamic phase transitions are expected with the corrections introduced by the GB sector and the higher dimensional framework. However, it is necessary to clarify that, in four dimensions in tensor-scalar theories with conformal symmetry \citep{Fernandes:2021dsb}, thermodynamic phase transitions have been found on an FLRW background \citep{Kong:2021dqd,Abdusattar:2023pck}. In the mentioned theory, the field equations have the same corrections as the Eqs. (\ref{FGB11}) and (\ref{FGB22}), but obtained in a four-dimensional setup.

\subsection{The equation of state}

The previous geometric framework and the equations of motion expressed in terms of the $AH$, are the first steps in going to a thermodynamic description in the cosmological scenario. Recalling that, here, one does not have an event horizon; instead, one has a dynamical entity, the $AH$, which is the pertinent surface to do thermodynamics in this context. Furthermore, the unified first law (UFL) should replace the thermodynamic first law. This was introduced in \citep{Hayward:1993wb,Hayward:1994bu,Hayward:1997jp}, in order to deal with dynamical BHs, where the event horizon is replaced by the $AH$ and the Killing vector fields by the Kodama vector fields \citep{Kodama:1979vn}. The thermodynamic UFL reads\footnote{It is worth mentioning that, in \citep{Maeda:2011ii}, \citep{Nozawa:2007vq} and \citep{Akbar:2006kj} was proven that the UFL also holds in the Lovelock, EGB and $f(R)$ gravity theories, respectively.}
\begin{equation}\label{UFL}
    \nabla_{i}E=A\psi_{i}+W\nabla_{i}V.
\end{equation}
In comparison with the usual first law employed in the gravitational context (for BHs) \citep{Padmanabhan:2002sha,Padmanabhan:2003gd,Wei:2009zzc}, here for the UFL (\ref{UFL}), the following replacements or extensions have been done \citep{Hayward:1997jp}: i) the gravitational active mass $M$ is recognized with Misner--Sharp (MS) energy\footnote{Although in the usual first law for BHs \citep{Padmanabhan:2002sha,Padmanabhan:2003gd,Wei:2009zzc}, this analogy is also established. That is, the mass $M$ is associated with the energy $E$ of the system. However, it should be emphasized that in this context, the energy $E$ is the MS energy and not other energy.} $E$ \citep{Misner:1964je}, ii) the term $TdS$ is replaced by the energy-supply, encoded by the vector $A\psi_{i}$. Finally, iii) the pressure $P$ of the system is identified as the density work $W$. 
These last two objects are defined by 

\begin{equation}\label{flux}
    \psi_{i}\equiv T^{j}_{i}\nabla_{j}R+W\nabla_{i}R,
\end{equation}
and 
\begin{equation}\label{workdensity}
    W\equiv-\frac{1}{2}h_{ij}T^{ij},
\end{equation}
respectively. The vector $\psi_{i}$ (\ref{flux}), is the so--called energy--flux.

As the thermodynamic analysis is done on the $AH$, one needs to project the UFL (\ref{UFL}) along it. So, projecting (\ref{UFL}) one obtains \citep{Hayward:1997jp} 
\begin{equation}\label{projected}  z^{i}\nabla_{i}E=\frac{\kappa_{\text{HK}}}{8\pi}z^{i}\nabla_{i}A+Wz^{i}\nabla_{i}V,
\end{equation}
where the following identification has been done $A\psi_{i}=\frac{\kappa_{\text{HK}}}{8\pi}z^{i}\nabla_{i}A$  \citep{Hayward:1997jp,Cai:2006rs} and $z^{i}$ is a vector tangent to the $AH$. 

In the projected UFL (\ref{projected}), the object $\kappa_{\text{HK}}$ is the so-called Hayward-Kodama (HK) surface gravity. As this is a purely geometric object (independent of the theory), it has the following definition \citep{Hayward:1997jp}
\begin{equation}\label{SG}
  \kappa_{\text{HK}}=\frac{1}{2\sqrt{-h}}\partial_{i}\left(\sqrt{-h}h^{ij}\partial_{j}R\right),  
\end{equation}
where $h\equiv \text{det}(h_{ij})$. Besides, the MS energy $E$ in this context, is being modified by the GB contributions, acquiring the following form \citep{Maeda:2006pm,Maeda:2007uu,Maeda:2008nz}
\begin{align}\label{MSEnergy}
E &\equiv \frac{(n-2)\pi^{(n-1)/2}}{8\pi G_{n}\Gamma\left[\frac{(n-1)}{2}\right]}\bigg[R^{n-3}\left(1-h^{ij}\nabla_{i}R\nabla_{j}R\right)+\tilde{\alpha} R^{n-5}\left(1-h^{ij}\nabla_{i}R\nabla_{j}R\right)^2\bigg].
\end{align}

The MS energy (\ref{MSEnergy}) evaluated at the FLRW metric (\ref{FLRW}) reads
\begin{equation}\label{energyFLRW}
    E_{\text{FLRW}}=\frac{(n-2)\pi^{(n-1)/2}}{8\pi G_{n}\Gamma\left[\frac{(n-1)}{2}\right]}\frac{R^{n-1}}{R^{2}_{AH}}\left[1+\frac{\tilde{\alpha}}{R^{2}_{AH}}\right].
\end{equation}
Of course, the above expression reduces to the GR case for $n=4$. The MS energy $E$ is not the only quantity receiving corrections from the GB sector, and the entropy $S$ does, too
\citep{Myers:1988ze,Cvetic:2001bk,Clunan:2004tb}
\begin{equation}\label{entropy2}
    S=\frac{A}{4G_{n}}\left[1+2\frac{\tilde{\alpha}}{R^{2}}\frac{(n-2)}{(n-4)}\right],
\end{equation}
with $A=2\pi^{(n-1)/2}R^{n-2}/\Gamma[(n-1)/2]$.

To perform the thermodynamic description, one needs an equation of state (EoS), that is, an equation relating the state variables $P=P(v,T)$. As stated in \citep{Hayward:1997jp}, the pressure $P$ yielding to the EoS in the context of the UFL is the work density (\ref{workdensity}). So we have
\begin{equation}\label{EoS}
\begin{split}
    W=P=\frac{(n-1)(n-2)}{16\pi R^{2}_{AH}}\left(1+\frac{\tilde{\alpha}}{R^{2}_{AH}}\right)
    +\frac{(n-2)}{16\pi}\left(1+\frac{\tilde{2\alpha}}{R^{2}_{AH}}\right)\left(\dot{H}-\frac{k}{a^{2}}\right).
    \end{split}
\end{equation}
Using (\ref{AHvalue}) to replace the last parenthesis in the right member of (\ref{EoS}) in favor of $\dot{R}_{AH}$ one gets
\begin{equation}\label{EoS1}
\begin{split}
  P=\frac{(n-1)(n-2)}{16\pi R^{2}_{AH}}\left(1+\frac{\tilde{\alpha}}{R^{2}_{AH}}\right)
    -\frac{(n-2)}{16\pi}\left(1+\frac{\tilde{2\alpha}}{R^{2}_{AH}}\right)\frac{\dot{R}_{AH}}{HR^{3}_{AH}}.
    \end{split}
\end{equation}
To obtain an expression of the form $P=P(V,T)$, it is convenient to introduce the specific volume\footnote{Following \citep{Kubiznak:2012wp}, the specific volume is defined as $v=2l^{2}_{p}R_{AH}$, where $l_{p}=\sqrt{G\hbar/c^{3}}$ is the Planck's length. Nevertheless, here we are assuming units where $\hbar=c=1$, and from now on, we will set $G_{n}=1$ for simplicity.} $v=2R_{AH}$, where the physical volume is given by $V=4\pi R^{3}_{AH}/3$. Moreover, the term containing $\dot{R}_{AH}$ can be eliminated using the expression of the surface gravity (\ref{SG}) on the FLRW (\ref{FLRW}) metric at the $AH$. Concretely, this is given by 
\begin{equation}\label{SGFLRW}
    \kappa_{\text{HK}}=\frac{1}{ R_{AH}}\left(\frac{\dot{R}_{AH}}{2HR_{AH}}-1\right).
\end{equation}

As it is known, there is a correlation between the surface gravity and the temperature of the system\footnote{See \citep{Cai:2008gw} for a related discussion about Hawking's radiation and temperature at the $AH$ of an FLRW cosmological model.} \citep{Hawking:1975vcx,Gibbons:1976ue}
\begin{equation}\label{temperature}
    T\equiv \frac{\kappa_{\text{HK}}}{2\pi},
\end{equation}
where the surface gravity has been replaced by the HK surface gravity. In the BHs context, the temperature (\ref{temperature}) should be positively defined. In the cosmological context, the situation changes radically (at least from the theoretical point of view). This means that both positive and negative temperatures are allowed. Obviously, the physical interpretation for positive and negative temperatures in the present scenario depends on the cosmological era and, consequently, the kind of matter filling the Universe. This point is quite relevant because after inserting (\ref{SGFLRW}) into (\ref{temperature}), one needs to take into account the sign of the right member of (\ref{SGFLRW}). However, as was discussed in \citep{Helou:2015yqa,Helou:2015zma}, the final sign of the temperature depends on future/past characteristics of the $AH$ combined with the resulting sign of the surface gravity, which depends on the inner/outer feature of the $AH$. So, before replacing $\dot{R}_{AH}$ in terms of the temperature $T$ into (\ref{EoS1}), to obtain the desired EoS, we are going to discuss the nature of the $AH$, matter distribution, surface gravity, and temperature.

Using the field equations (\ref{FGB11})--(\ref{FGB22}) the surface gravity (\ref{SGFLRW}) can be written as
\begin{equation}\label{SG2}
    \kappa_{\text{HK}}=-\frac{R_{AH}}{2}\left(2H^{2}+\dot{H}+\frac{k}{a^{2}}\right).
\end{equation}
Now, assuming that the thermodynamic variables of the energy-momentum tensor (\ref{EMT}) are linked by a barotropic EoS 
\begin{equation}
    p(\rho)=\omega \rho,
\end{equation}
with $\omega$ the EoS parameter. Then (\ref{SG2}) reads 
{
\begin{equation}\label{SGMatter}
    \kappa_{\text{HK}}=\frac{4\pi G_{n}\rho}{(n-2)}\frac{(\omega+1)(n-1)-4}{(n-1)}R_{AH}+\frac{\tilde{\alpha}}{R^{3}_{AH}}(1-\dot{R}_{AH}).
\end{equation}}

{As can be seen, the signature of the surface gravity depends on several factors: i) The EoS parameter, ii) the dimensions $n$ and iii) the GB corrections through the rate expansion/contraction of the Universe. In this concern, as the temperature is proportional to the surface gravity, that is, $T\propto \kappa_{\text{HK}}$, it is so important to determine whether $\kappa_{\text{HK}}$ is positive or negative in nature. Particularly, one has $T\propto +\kappa_{\text{HK}}$ for future $AH$ and $T\propto -\kappa_{\text{HK}}$ for past $AH$. The future or past character of the $AH$ depends on the cosmological scenario, that is, contracting or expanding, respectively. While the inner/outer feature concerns the sign of the surface gravity, being inner for negative and outer for positive surface gravity. In general, one has the following combinations \citep{Helou:2015yqa,Helou:2015zma}: i) future/outer, ii) past/inner, iii) past/outer, and iv) future/inner horizons. The $AH$ is past/inner for expanding cosmologies and future/inner for contracting cosmologies. So, the final sign of the temperature in the former is positive and negative in the second case. Besides these scenarios, there is another one in the category of expanding cosmologies, where the matter distribution is the so-called \emph{stiff} matter. In such a case, the barotropic EoS is given by $p(\rho)=\rho$, that is, with EoS parameter\footnote{In table \ref{table1}, we have placed for stiff matter the condition $\omega\geq 1$. Actually, the stiff matter takes the equality. However, in some speculative cosmological models, one can consider the EoS parameter greater than one.} $\omega=1$. This kind of matter has been used in the study of the early Universe stage, leading to a classical bouncing model, free from singularities \citep{Oliveira-Neto:2011uhf} and also to describe the same problem from the quantum point of view \citep{Falciano:2007yf}.

Now, if one takes into account the usual matter distributions whose have dominated the different epochs, that is, inflation ($\omega=-1$), dust ($\omega=0$), dark energy ($\omega=-1$) and radiation ($\omega=1/3$), all of them correspond to inner/past $AH$ with a positive temperature. However, in the GR case, the radiation epoch has a degenerate $AH$, that is, an $AH$ where $\kappa_{\text{HK}}=0$ consequently $T=0$ \citep{Helou:2015yqa,Helou:2015zma}. Nevertheless, when GB corrections are present, this is no longer the case. Furthermore, for the inflation and cosmological constant scenario, one needs to demand $\dot{R}_{AH}=0$, that is, a quasi-adiabatic process. On the other hand, dust epoch and radiation entail $\dot{R}_{AH}>1$ and $\dot{R}_{AH}>>1$, respectively. The era of hypothetic stiff matter imposes $\dot{R}_{AH}>0$ leading to $T<0$. Finally, in the event of a contracting scenario due to the distribution of phantom matter, the $AH$ is inner/future in nature with $T<0$. For this to happen, one needs $\dot{R}_{AH}<<0$.}

In summary, the type of matter distribution present in different epochs of the Universe's evolution allows us to rapidly determine the sign of the surface gravity and then the temperature. Actually, this conclusion comes from $\gamma\equiv \mathcal{L}_{-}\theta_{-}/\mathcal{L}_{+}\theta_{-}$, that is, the ratio of Lie derivatives of the expansion coefficients. This determines the causal nature of a horizon \citep{Dreyer:2002mx}. Table  \ref{table1} shows some of the possibilities given in the GB context.

\begin{table}[H]
\resizebox{1.001 \hsize}{!}{$
\begin{tabular}{|c|c|c|c|c|}
\hline
Cosmological era                             &  Expanding/Contracting   & Inner/Outer                    & Future/Past & Temperature              \\ \hline
Inflation ($\omega=-1$)                      &  Expansion ($\dot{R}_{AH}=0$)   & Inner ($\kappa_{\text{HK}}<0$) & Past        & $T\propto -\kappa_{\text{HK}} \Rightarrow T>0$ \\ \hline
Dust ($\omega=0$)                          &  Expansion ($\dot{R}_{AH}>0$)   & Inner ($\kappa_{\text{HK}}<0$) & Past        & $T\propto -\kappa_{\text{HK}} \Rightarrow T>0$ \\ \hline
Radiation ($\omega=1/3$)                     &  Expansion ($\dot{R}_{AH}>0$)   & Inner ($\kappa_{\text{HK}}<0$) & Past        & $T\propto -\kappa_{\text{HK}}$ $\Rightarrow T>0$ \\ \hline
Cosmological constant  ($\omega=-1$)         &  Expansion ($\dot{R}_{AH}=0$)      & Inner ($\kappa_{\text{HK}}<0$) & Past        & $T\propto -\kappa_{\text{HK}} \Rightarrow T>0$ \\ \hline
Stiff matter ($\omega\geq1$)                    &  Expansion ($\dot{R}_{AH}>0$)   & Outer ($\kappa_{\text{HK}}>0$) & Past        & $T\propto -\kappa_{\text{HK}} \Rightarrow T<0$ \\ \hline
Phantom ($\omega\textless{}-1$) &  Contracting ($\dot{R}_{AH}<0$) & Inner ($\kappa_{\text{HK}}<0$) & Future      & $T\propto +\kappa_{\text{HK}} \Rightarrow T<0$ \\ \hline
\end{tabular}$}
\parbox{\textwidth}{\caption{The  temperature expression (last column) for different types of cosmological eras. }\label{table1}}
\end{table}

Now, the above simple analysis allows us to discriminate the expression for the temperature (last column in table \ref{table1}) to be inserted in Eq. (\ref{EoS1}) according to the Universe era. Nevertheless, it should be taken into account that this analysis was done in order to give an interpretation of possible thermodynamic phenomena depending on the universe's era. So, for a past/inner and past/outer $AH$, the EoS (\ref{EoS1}) becomes

\begin{equation}\label{eosip}
    P(v,T)=\frac{(n-2)T}{2v}+\frac{(n-2)(n-3)}{4\pi v^{2}}+\frac{4\tilde{\alpha} (n-2)T}{v^{3}}+\frac{\tilde{\alpha}(n-2)(n-5)}{\pi v^{4}},
\end{equation}
and for future/inner, the EoS is
\begin{equation}\label{eosif}
    P(v,T)=-\frac{(n-2)T}{2v}+\frac{(n-2)(n-3)}{4\pi v^{2}}-\frac{4\tilde{\alpha} (n-2)T}{v^{3}}+\frac{\tilde{\alpha}(n-2)(n-5)}{\pi v^{4}}.
\end{equation}
As can be seen, the main difference between (\ref{eosip}) and (\ref{eosif}) is the minus sign in from those terms involving the temperature $T$. This simple difference could drastically change the thermodynamic behavior and determination of the existence of critical phenomena. This point shall be a matter of fact in the following section. 

\section{P--V phase transitions }\label{sec3}

To explore the possibility of having $P-v$ phase transitions, the following conditions are required 
\begin{equation}\label{criticalpoints}
    \left(\frac{\partial P(v,T)}{\partial v}\right)\bigg{|}_{T}=0, \quad \left(\frac{\partial^{2}
    P(v,T)}{\partial v^{2}}\right)\bigg{|}_{T}=0.
\end{equation}

Solving the above conditions for the EoS (\ref{eosip}), one gets the following critical points for the reduced volume $v_{c}$ and temperature $T_{c}$

\begin{equation}\label{rvolume}
    v_{c1}=2\sqrt{6(n-4)\alpha-2\sqrt{3}\sqrt{(6-n)(n-4)^{2}(n-2)\alpha^{2}}},
\end{equation}
\begin{equation}\label{rtemperature}
    T_{c1}=-\frac{\sqrt{3(n-4)\alpha-\sqrt{3}\sqrt{(6-n)(n-4)^{2}(n-2)\alpha^{2}}}\left((n-4)n \alpha+\sqrt{3}\sqrt{(6-n)(n-4)^{2}(n-2)\alpha^{2}}\right)}{12\sqrt{2}(n-4)^{2}(n-3)\pi \alpha^{2}},
\end{equation}
\begin{equation}\label{rvolume1}
    v_{c2}=2\sqrt{6(n-4)\alpha+2\sqrt{3}\sqrt{(6-n)(n-4)^{2}(n-2)\alpha^{2}}},
\end{equation}
\begin{equation}\label{rtemperature1}
    T_{c2}=\frac{\sqrt{3(n-4)\alpha+\sqrt{3}\sqrt{(6-n)(n-4)^{2}(n-2)\alpha^{2}}}\left((n-4)n \alpha+\sqrt{3}\sqrt{(6-n)(n-4)^{2}(n-2)\alpha^{2}}\right)}{12\sqrt{2}(n-4)^{2}(n-3)\pi \alpha^{2}}.
\end{equation}

Notice that we have replaced  $\tilde{\alpha}$ in the above expressions. For the EoS (\ref{eosif}), there are no critical points satisfying the corresponding conditions about the Universe era. In this case, one obtains a positive temperature instead of a negative one. So, EGB theory has no phase transition for a contracting cosmological scenario.

Before computing the critical pressure $P_{c}$ for solutions (\ref{rvolume})-(\ref{rtemperature}) and (\ref{rvolume1})-(\ref{rtemperature1}), it is important to analyze under what conditions the critical volumes and the critical temperatures meet the analysis given in table \ref{table1}. 
From expression (\ref{rvolume}), it is clear that the space-time dimension $n$ must be less than or equal to six to avoid complex values and provides positive volume, in this case $5<n\leq 6$. In addition, the coupling $\alpha$ should also be positive. On the other hand, for (\ref{rtemperature}) it is not possible to get a positive temperature. Therefore, inflation, dust and cosmological constant eras are ruled out. In this case, only negative temperatures subject to $5\leq n\leq 6$ are possible. It is clear that the critical volume imposes more restrictions with respect to space-time dimensions. Now, for (\ref{rvolume1}) and (\ref{rtemperature1}), the conditions are $5\leq n\leq 6$ for positive critical volume and negative critical temperature. This case is more involved since it allows us to incorporate five-dimensional space-times. So, from now on we are going to use the pair $\{v_{c2},T_{c2}\}$, since the first pair is included in this case (only the six-dimensional case). In summary, the present scenario corresponds to a hypothetical Universe dominated by stiff matter or matter with an EoS parameter greater than one. 

It is worth mentioning that the four-dimensional case and lower ones are automatically discarded by the bound given above. This supports the fact that the GB term (\ref{GBinvariant}) becomes a topological invariant in four dimensions and it is not given any contribution to lower dimensions than four. So, it is clear the preponderant role played by the space-time dimension $n$ in the existence of phase transitions, where above $n=6$ dimensions, there are no critical points (real roots with physical interpretation). Hence phase transitions do not occur in dimensions greater than six.

Now, the critical pressure $P_{c2}$ is obtained by replacing (\ref{rvolume1}) and (\ref{rtemperature1}) into (\ref{eosip}), yielding to\footnote{From now on, as we are going to work with $\{P_{c2},v_{c2},T_{c2}\}$ only, we are going to drop out the sub-script 2 to mention the critical values.}
\begin{equation}
    P_{c2}=\frac{(n-2)(n-3)(n-4)\left((4-n)(3-13n+2n^{2}) \alpha+\sqrt{3}(n-9)\sqrt{(6-n)(n-4)^{2}(n-2)\alpha^{2}}\right)\alpha}{64\pi\left(3(n-4) \alpha+\sqrt{3}\sqrt{(6-n)(n-4)^{2}(n-2)\alpha^{2}}\right)^{3}}.
\end{equation}

As this stage, we discuss the critical behavior of (\ref{eosip}), that is, the possible thermodynamic phase transitions for cases $n=5$ and $n=6$ \\

\begin{figure}[H]
    \centering
    \includegraphics[width=0.42\textwidth]{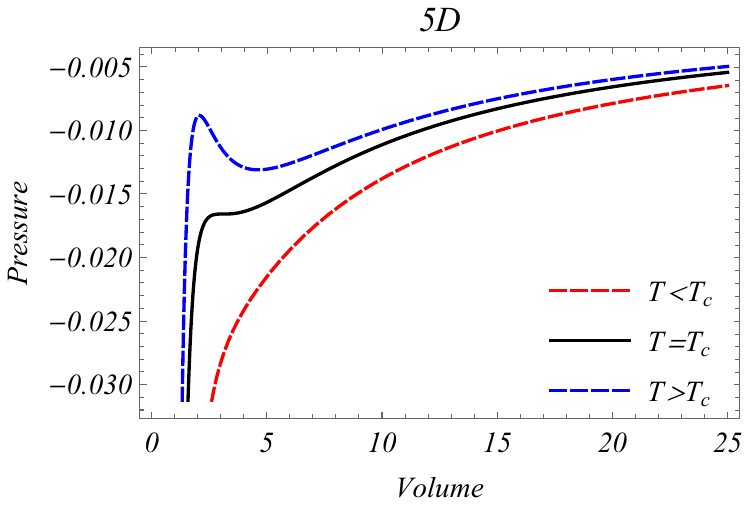}\   \includegraphics[width=0.42\textwidth]{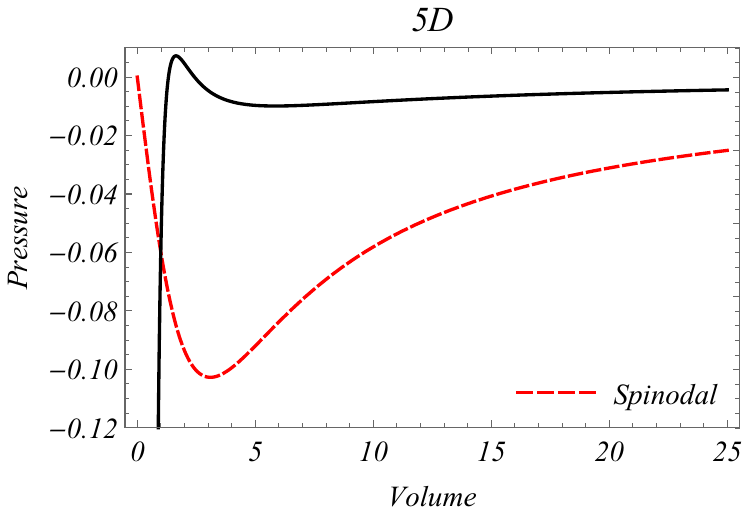} 
    \caption{\textbf{Left panel}: The behavior of the pressure (\ref{eosip}) against the molar volume for different values of the  temperature. As can be observed, for those values above the critical one, the system is exhibiting an inverted first order phase transition (dashed blue line), where the \emph{wiggle} now corresponds to stable states and the remaining one to unstable states. On the other hand, for temperature below the critical one, the system is completely unstable (dashed red line). \textbf{Right panel}: The pressure and the ``spinodal'' curve versus the volume. For these plots we fixed $\alpha=0.2$ square length. }
    \label{fig1}
\end{figure}

\begin{figure}[H]
    \centering
    \includegraphics[width=0.42\textwidth]{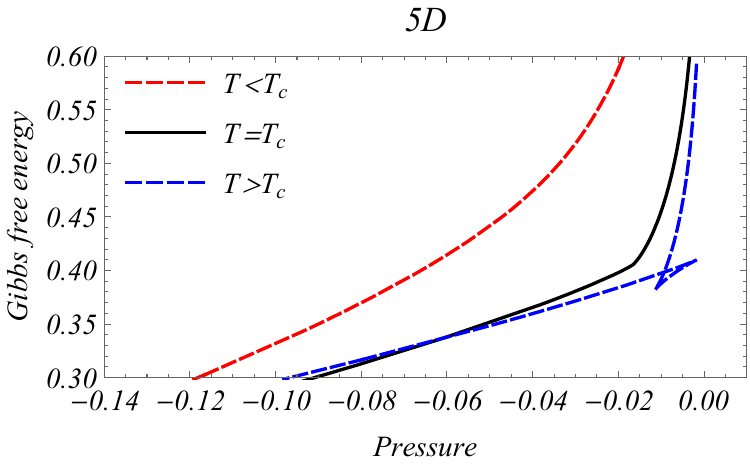}\   \includegraphics[width=0.42\textwidth]{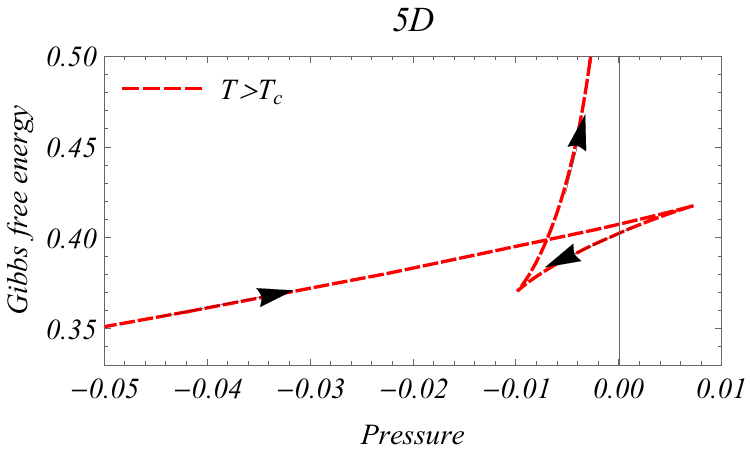} \\
\includegraphics[width=0.42\textwidth]{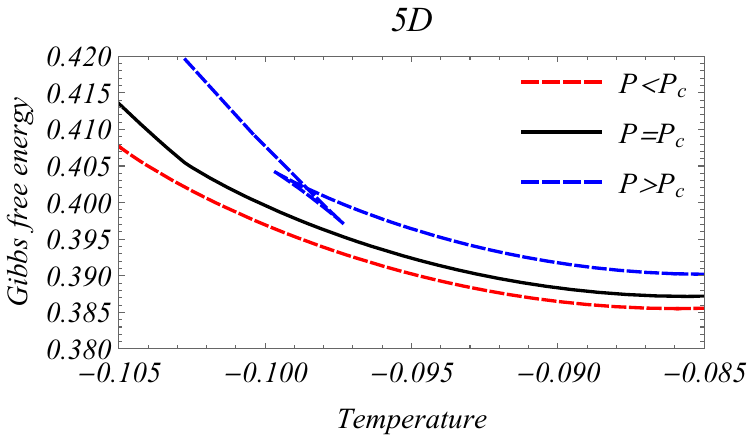}\   \includegraphics[width=0.42\textwidth]{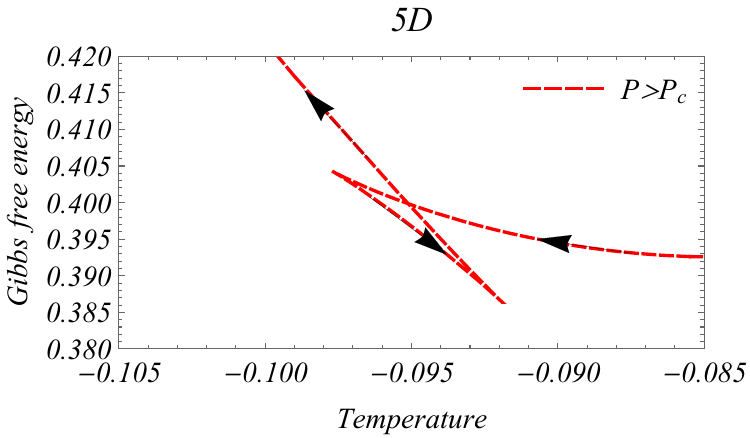}
    \caption{\textbf{Top row}: The behavior of the Gibbs free energy versus the pressure for different values of the temperature. \textbf{Bottom row}: The behavior of the Gibbs free energy versus the temperature for different values of the pressure. The Gibbs free energy in this case displays a non-physical swallow-tail for $T>T_{c}$ (see the direction given by arrows for right panels on the top and bottom row). For these plots we fixed $\alpha=0.2$ square length.   }
    \label{fig1-1}
\end{figure}

$\textbf{n=5}$ \textbf{dimensions}:

\begin{itemize}
    \item The left panel in Fig. \ref{fig1} is displaying the  pressure (\ref{eosip}) trend versus the molar volume. This system shows an unconventional behavior with respect to those systems that undergo liquid-gas-like phase transitions. Here, the phase transformation occurs for temperature above the critical one (above the black line), while below the critical temperature (dashed red line), the system does not exhibit any drastic change or behavior. In this case, the dashed blue line depicts an ``inverted'' first order phase transition (inverted Van der Waals system). By Inverted we mean regions corresponding to stable states in the usual Van der Waals system, now are unstable and vice versa. 
    \item The right panel of Fig. \ref{fig1}, is depicting the behavior of the pressure for temperatures above the critical one (black line) and the ``spinodal'' curve (dashed red line). In this case, we cannot say that it is in fact the spinodal curve as happens in the Van der Waals system. The reasons are clear, by definition the spinodal curve marks the boundary where a homogeneous fluid becomes mechanically unstable with respect to phase separation. This curve represents the limit of metastability, beyond which the system spontaneously separates into two distinct phases (liquid and vapor) without the need for fluctuations to nucleate the phases. However, in this case this definition does not apply. Here we have a confined stable region, then we cannot mark or stablish the limit between these states using the spinodal curve. 
    
    \item Fig. \ref{fig1-1} Shows the behavior of the Gibbs free energy against the pressure (top row) and versus the temperature (bottom row).  As can be seen, there is a loop (swallowtail-like) portion representing in this case stable  states, that is, regions where $\partial{{P}}/\partial{v}<0$. However, the remaining portions represent the unstable states. This is so because these states are not in thermodynamic equilibrium. This is contrary to what happens 
for liquid-gas phase transition, as in van der Waals's real gas case. In this case, we can say that the swallow-tail is not physical, because it does not minimize the free energy. It should be noted that this loop is formed for those temperature above the critical one. 
\end{itemize}

\begin{figure}[H]
    \centering
    \includegraphics[width=0.42\textwidth]{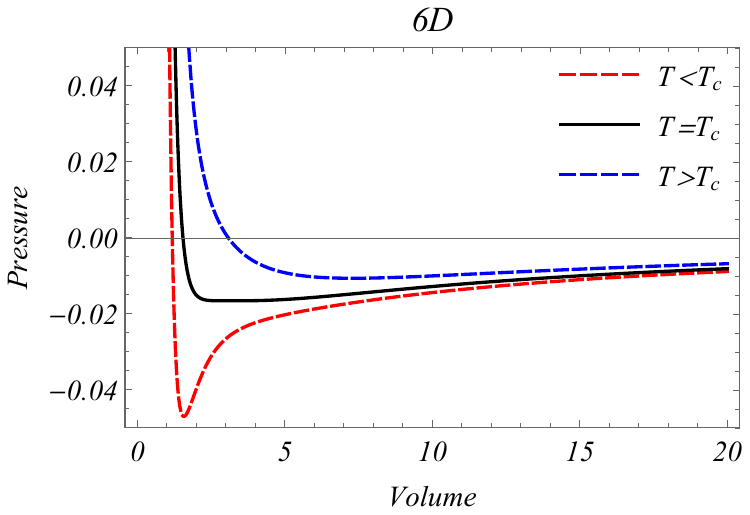}\       \includegraphics[width=7.2cm,height=5.1cm]{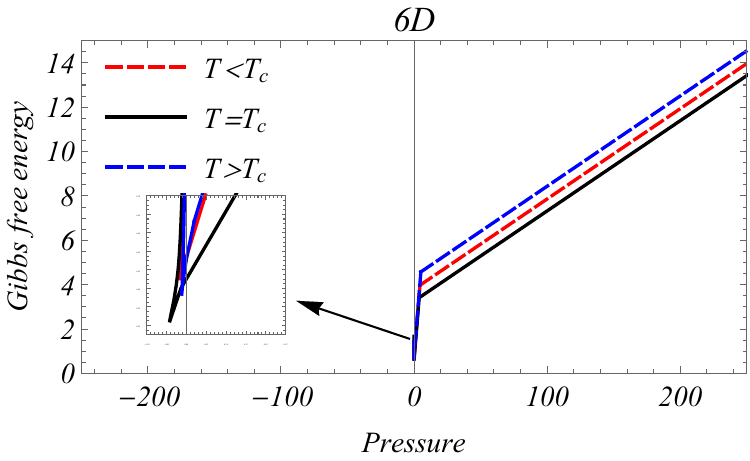}
    \caption{\textbf{Left panel}: The trend of the pressure the molar volume. \textbf{Right panel}: The trend of the Gibb free energy against the pressure for different values of the temperature. For these plots we fixed $\alpha=0.2$ square length.}
    \label{fig2}
\end{figure}

$\textbf{n=6}$ \textbf{dimensions}:

\begin{itemize}
    \item The left and right panels of Fig. \ref{fig2}, is displaying the behavior of the  pressure against the molar volume. Here the system does not present a phase transition above the critical temperature. In this case, the system has stable/unstable branches. For small volumes, the system is stable but after some point it becomes unstable.  
    \item The Gibbs free energy is depicted in Fig. \ref{fig2} (right panel), for temperatures below and above the critical one. Here the Gibbs free energy shows its characteristic cuspy behavior, that is, it is a continuous function. Notwithstanding, as the unstable states dominates, the free energy increases with increasing pressure (volume), then the system does not tend to the thermodynamic equilibrium. 
\end{itemize}

At this point, it is important to comment about the first results we obtained. As can be appreciated, from the mathematical point of view there is a natural constraint on the admissible dimensions providing thermodynamics phase transitions. In this case the system yields to an upper bound for the space-time dimensions, being it $n=6$ and a lower one given by $n=5$. The minimum admissible dimension seems to be a reasonable constraint since for dimensions lower than five, the GB terms is not affecting the dynamic of the theory. The upper space-time dimension bound, in this case appears from a pure mathematical limit preventing the system to have only real solutions for both, the critical molar volume and critical temperature.
On the other hand, for $n=5$ and $n=6$, the thermodynamics behavior is quite different. This can be related with the contribution given the GB term in five and six dimensions, respectively. In this case we can argue that The differences in phase transition behavior between $n=5$ and $n=6$ EGB theory arise due to the interplay of dimensionality, topology, and dynamics. In five dimensions, the system is constrained, and curvature corrections dominate, leading to an inverted first-order phase transition. In contrast, in six dimensions, the greater degrees of freedom and topological complexity result in a smoother thermodynamic evolution, preventing phase transitions. The nature of the GB term influence diminishes as the number of dimensions increases, contributing to the absence of sharp thermodynamic phase transitions in six dimensions.

\section{Geometrothermodynamics and Critical phenomena}\label{sec4}

In order to get a more precise description of the thermodynamics behavior and associated properties of the FLRW space-time in the EGB framework, it is convenient to do an analysis of the microscopic structure throughout the so-called Riemannian thermodynamics developed by Ruppeiner \citep{Ruppeiner:1981znl,Ruppeiner:1983zz,Ruppeiner:1995zz}. 

This geometric description of the thermodynamic fluctuations provides a powerful diagnostic for the microstructures of astrophysical/cosmological models. Basically, 
through the examination of the thermodynamic curvature $R$ (this is linked to microstructure interactions), it is possible to understand the
microstructure of the systems better. The main ingredient is the sign of thermodynamic curvature $R$, which determines whether the microstructure interactions are repulsive or attractive in nature, that is when $R>0$
repulsive interactions prevail, and for $R < 0$ attractive interactions are present. The critical point $R=0$ means no interactions are present.  

\subsection{Ruppeiner Geometry}
The starting point is line element $dl^{2}$ expressed in terms of thermodynamic variables as\footnote{For the microstructure analysis of black holes solutions applying Ruppeiner's procedure, in the GR, EGB and beyond see for instance \citep{Wei:2015iwa,Wei:2019uqg,Wei:2019yvs,Wei:2019ctz,Ruppeiner:1983zz,Ruppeiner:1995zz,Ruppeiner:1981znl,Ruppeiner:2013yca,Ruppeiner:2018pgn,Ruppeiner:2023wkq}.}
\begin{equation}\label{ruppe}   dl^{2}=g_{\mu\nu}dx^{\mu}dx^{\nu},
\end{equation}
where the coordinates $x^{\mu}$ are functions of the internal energy $U$ and molar volume $v$ and the metric $g_{\mu\nu}$ is given in terms of the entropy $S$ as follows
\begin{equation}
g_{\mu\nu}=-\frac{\partial^{2}S}{\partial x^{\mu}\partial x^{\nu}}.
\end{equation}
After some manipulation using thermodynamics relations, the line element can be cast as \citep{Wei:2015iwa}
\begin{equation}\label{rupeline}
d l^2=\frac{1}{T}\left(-\frac{C_v}{T} d T^2+\left(\frac{\partial P}{\partial v}\right)\bigg|_T d v^2\right),
\end{equation}
being $C_{v}$ the heat capacity at constant volume $v$. As for this system $T<0$, it is important to clarify that the line element (\ref{rupeline}) is not changing its signature. This is so because the first term is quadratic in $T$ and the second term naturally incorporates the change in sign through 
$\frac{\partial P}{\partial v}$. Since, for thermodynamics system there is a possibility of having $\frac{\partial P}{\partial v}<0$ or $\frac{\partial P}{\partial v}>0$. Another important point to be highlighted is the divergence occurred at $c_{v}=0$. This happens because
the (\ref{eosip}) is linear in $T$, like in the van der Waals model, then heat capacity at constant molar volume $C_{v}$ is vanishing. In this way, the line element (\ref{ruppe}) becomes singular. Therefore, the associated thermodynamic curvature is divergent. 
To check this point, from (\ref{eosip}) and (\ref{ruppe}) it is not hard to obtain
\begin{equation}\label{scalarR}
    R=\frac{\left[\left(n-3\right)v^{2}+8\left(n-5\right)\tilde{\alpha}\right]\left[\left(n-3\right)v^{2}-2\pi v^{3}T+8\left(n-5\right)\tilde{\alpha}-48\pi \tilde{\alpha} vT\right]}{2C_{v}\left[\left(n-3\right)v^{2}-\pi v^{3}T+8\left(n-5\right)\tilde{\alpha}-24\pi\tilde{\alpha} vT\right]^{2}}.
\end{equation}

As can seen from (\ref{scalarR}), when $C_{v}=0$, the thermodynamic scalar curvature blows up. To bypass this issue, in \citep{Wei:2019uqg,Wei:2019yvs,Wei:2019ctz} was proposed the normalized thermodynamic scalar curvature $R_{N}$ 

\begin{equation}\label{Recurvature}
R_{\mathrm{N}}=C_{v}R=\frac{\left[\left(n-3\right)v^{2}+8\left(n-5\right)\tilde{\alpha}\right]\left[\left(n-3\right)v^{2}-2\pi v^{3}T+8\left(n-5\right)\tilde{\alpha}-48\pi \tilde{\alpha} vT\right]}{2\left[\left(n-3\right)v^{2}-\pi v^{3}T+8\left(n-5\right)\tilde{\alpha}-24\pi\tilde{\alpha} vT\right]^{2}}.
\end{equation}

\begin{figure}[H]
    \centering
    \includegraphics[width=0.42\textwidth]{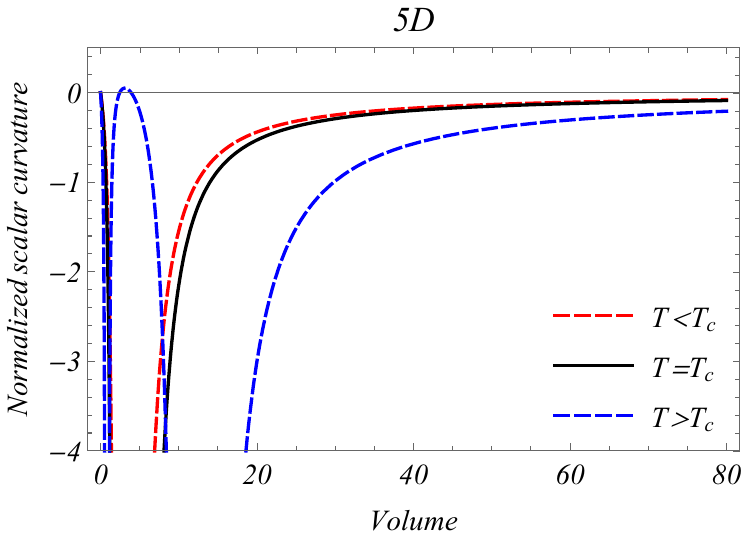}\   \includegraphics[width=0.42\textwidth]{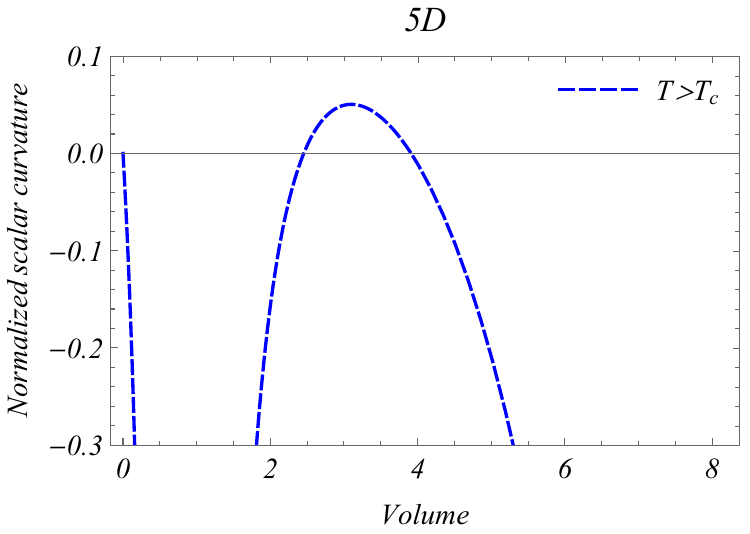} \\
\includegraphics[width=0.42\textwidth]{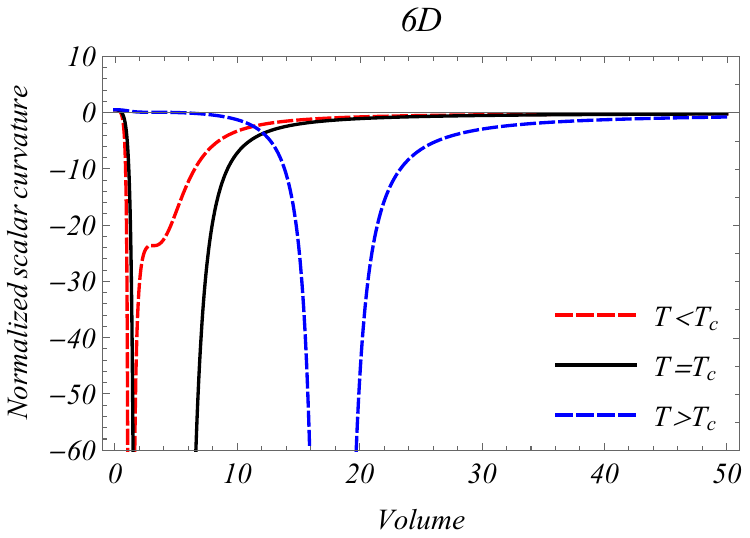}\   \includegraphics[width=0.42\textwidth]{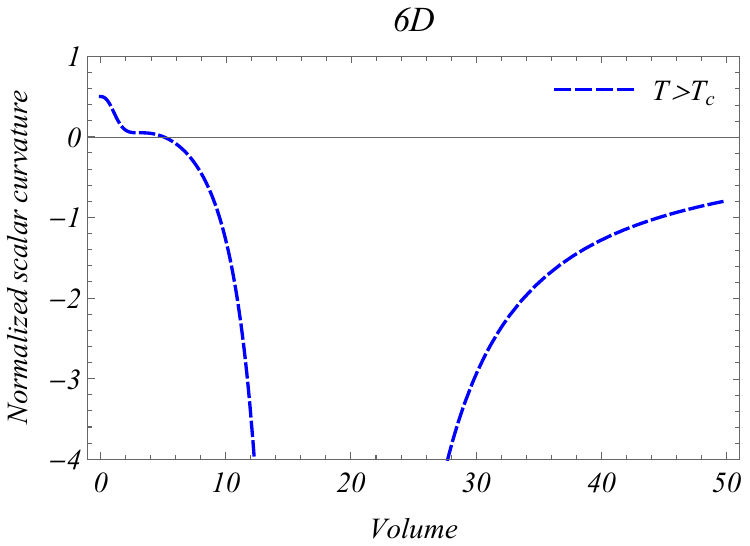}
    \caption{\textbf{Top row}: The normalized scalar curvature versus the  volume. The \textbf{left} panel is showing the normalized curvature trend different temperatures. For some values, the scalar curvature is negative, positive, and again negative. This fluctuation reveals the presence of a phase transition. Points, where the scalar curvature nullifies, correspond to those values of ${T}$ satisfying ${R}_{N}=0$, giving rise to the so-called sign-changing curve. The right panel shows a zoom of the normalized curvature when it changes in sign for $T>T_{c}$. \textbf{Bottom row}: In this case, the normalized scalar curvature trend is quite different from the previous situation. Here, this invariant starts from a positive position for $T>T_{c}$ and then becomes negative in nature. This confirms that for $n=6$ no phase transitions are occurring. Besides, repulsive interactions (unstable states) and attractive interactions (stable states) exist. For these plots we fixed $\alpha=0.2$ square length.}
    \label{fig4}
\end{figure}

The trend of the normalized scalar curvature (\ref{Recurvature}) against the molar volume is depicted in Fig. \ref{fig4}. The top panels correspond to $n=5$, where the left one shows the behavior of ${R}_{N}$ for different temperatures. For $T>T_{c}$ (dashed blue line), the normalized curvature for small volumes is negative, positive and negative again (see the zoom on the right panel in Fig. \ref{fig4}). This means that there is a change in sign reflecting a phase transition. Here the interpretation is different from the given for BHs \cite{Wei:2019yvs}, namely, the region $R_{N}>0$ in this case corresponds to attractive interactions while $R_{N}<0$ repulsive interactions. So, the system goes from repulsive interactions (unstable states), attractive interactions (stable states) and repulsive interactions again. For $T<T_{c}$ (dashed red line) the normalized curvature is always negative, this means that the system is undergoing repulsive interactions among the particles. This belongs to the unstable states. Notably in both cases, $T>T_{c}$ and $T<T_{c}$, $R_{N}$ has two divergent points merging at $T=T_{c}$ (solid black line).

In the case $n=6$ (lower panels in Fig. \ref{fig4}), the right panel shows the trend of the normalized curvature for different temperatures. In this case also there are two divergent points for $T>T_{c}$ and $T<T_{c}$ merging at $T=T_{c}$. In comparison with the $n=5$ case, here there is only one change in sign, from positive (for small volumes) to negative. Positive curvature here means attractive interactions in the stable zone, while negative curvature repulsive interactions corresponding o unstable states as anticipated.

\begin{figure}[H]
    \centering
    \includegraphics[width=0.42\textwidth]{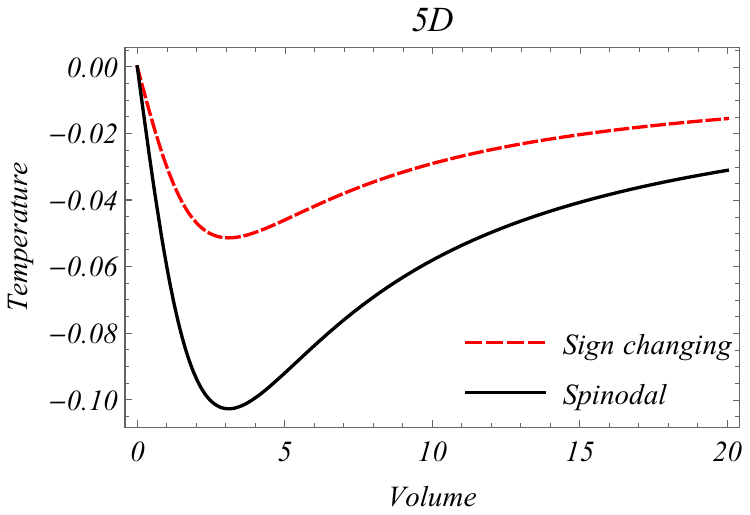}\     \includegraphics[width=0.42\textwidth]{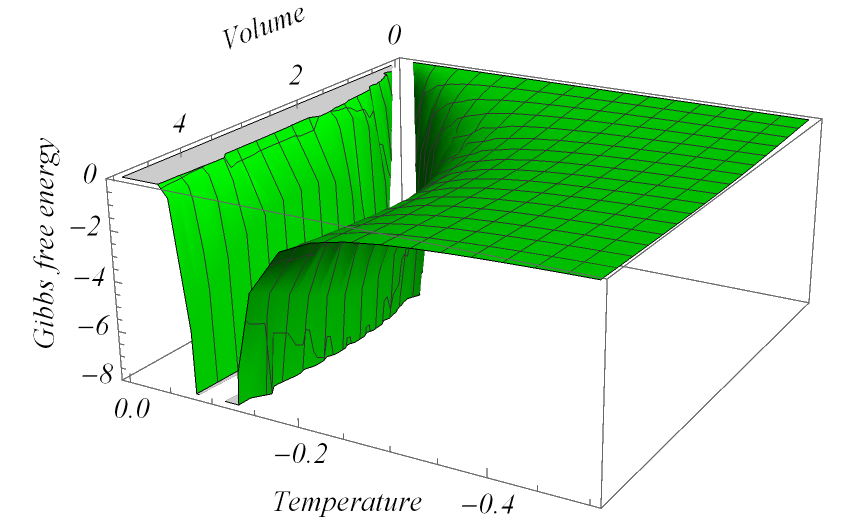} \\
\includegraphics[width=0.42\textwidth]{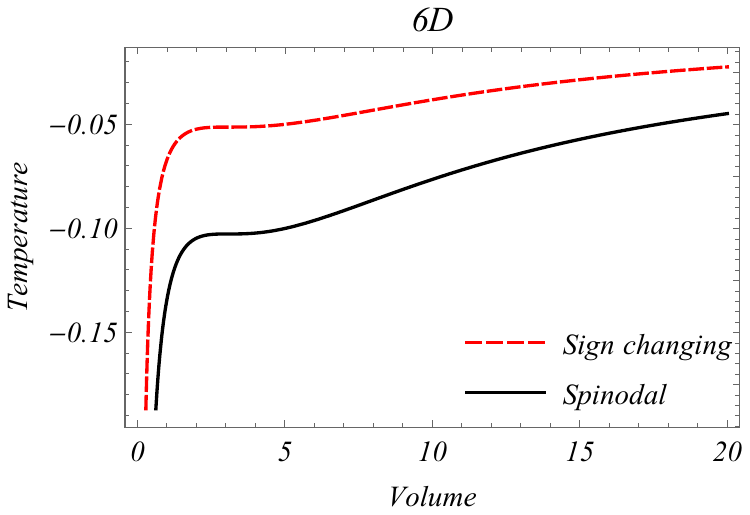}\     \includegraphics[width=0.42\textwidth]{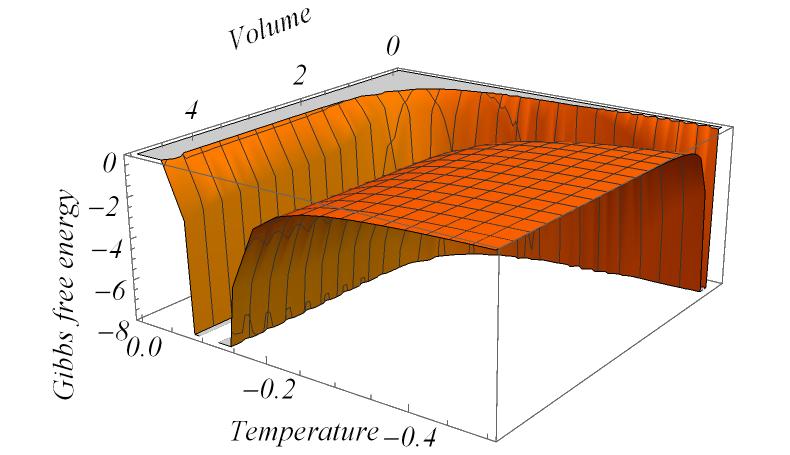}
    \caption{\textbf{Top row}: The left panel shows the sign changing curve (dashed red line) and the ``spinodal'' curve (black line). The right panel displays the 3D normalized curvature behavior. These plots correspond to $n=5$ \textbf{Bottom row}: The same as the top row, but this time for $n=6$. For these plots we fixed $\alpha=0.2$ square length.  }
    \label{fig5}
\end{figure}

The change in the sign of the scalar curvature occurs at a particular curve, the so-called sign-changing curve. This information can be obtained by solving ${R}_{N}=0$, leading to 
\begin{equation}\label{zeron5}
{T}_{0}=\frac{[3-n][v^2+8(n-5)(n-4)\alpha]}{2\pi v[v^2+24(n-4)(n-3)\alpha]},
\end{equation}
On the other hand, the  normalized scalar curvature (\ref{Recurvature}) clearly diverges at the following reduced temperature 
\begin{equation}\label{div1}
{T}_{\text{div}}=\frac{[3-n][v^2+8(n-5)(n-4)\alpha]}{\pi v[v^2+24(n-4)(n-3)\alpha]},
\end{equation}
notice that in Eqs. (\ref{zeron5}) and (\ref{div1}) we have replaced 
$\tilde{\alpha}$ to get a better visualization of these curves.
Interestingly, the curve at which ${R}_{N}=0$, is related to the divergence curve ${T}_{\text{div}}$, via ${T}_{0}={T}_{\text{div}}/2$. Furthermore, this latter coincides with the limit beyond which a mixture becomes unstable and undergoes spontaneous phase separation (in case of a genuine first order phase transition), that is, the ``spinodal'' curve. Therefore, ${T}_{\text{div}}={T}_{\text{spin}}$ and 
\begin{equation}
    \lim_{{T}\rightarrow {T}_{\text{spin}}}{R}_{N}\rightarrow -\infty.
\end{equation}

As said before, the cannot talk about spinodal curve, however, we keep this name in our description, but in this case this curve has not the physical meaning that it has for a genuine first order phase transition. What is more, for those system not undergoing first order phase transitions, the spinodal curve does not exist. This is clear from the lower left panel in Fig. \ref{fig5} for the case $n=6$. For the case $n=5$, the sign changing and spinodal curves coincide for small values of the molar volume, then they approach asymptotically. That's why we cannot say that this a real or physical spinodal curve because it is not limiting the phases.

\subsection{Critical phenomena}

In principle, critical exponents are primarily associated with second-order phase transitions rather than first-order phase transitions \citep{Goldenfeld:1992qy}. However, critical exponents can still describe the behavior near the critical point for first-order phase transitions. These exponents provide insight into the nature of phase transitions and the behavior of thermodynamic properties in the vicinity of critical points, even for systems undergoing first-order phase transitions like the van der Waals gas, for example.

In general, to determine the critical exponents $\{\hat{\alpha},\hat{\beta},\hat{\gamma},\hat{\delta}\}$ it uses the approximate behavior near the critical point of the heat capacity at constant volume $C_{v}$, the shear viscosity $\eta$, the compressibility $\kappa_{T}$ and pressure $P$, as follows
\begin{equation}\label{exponent}
\begin{aligned}
& C_v \sim|\tau|^{-\hat{\alpha}}, \\
& \eta \sim v_l-v_s \sim|\tau|^{\hat{\beta}}, \\
& \kappa_T=-\frac{1}{v}\left(\frac{\partial v}{\partial P}\right)\bigg{|}_T \sim|\tau|^{-\hat{\gamma}}, \\
& P-P_c \sim\left|v-v_c\right|^{\hat{\delta}}.
\end{aligned}
\end{equation}
In the above expressions, $v_{l}$ and $v_{s}$ stand for the large and small volumes, or in terms of normal thermodynamics, $l$
and $s$ may be vapour and liquid phases and $\tau\equiv\frac{T}{T_{c}}-1$.

In the present case, as $C_{v}=0$, then $\hat{\alpha}=0$, as for the van der Waals gas. Next, to find out the other critical exponents, we expand the pressure $P$ around the critical values $v_{c}$ and $T_{c}$, yielding to

\begin{equation}\label{expansion}
\begin{aligned}
P= & P_c+\left[\left(\frac{\partial P}{\partial T}\right)\bigg|_v\right]\bigg|_c\left(T-T_c\right) +\frac{1}{2 !}\left[\left(\frac{\partial^2 P}{\partial T^2}\right)\bigg|_v\right]\bigg|_c\left(T-T_c\right)^2  +\left[\left(\frac{\partial^2 P}{\partial T \partial v}\right)\right]\bigg|_c\left(T-T_c\right)\left(v-v_c\right) \\
& +\frac{1}{3 !}\left[\left(\frac{\partial^3 P}{\partial v^3}\right)\bigg|_T\right]\bigg|_c\left(v-v_c\right)^3+\ldots
\end{aligned}
\end{equation}
In the above expression, we have used the fact that at the critical point $(\partial P/\partial v)_{c}=(\partial^{2} P/\partial v^{2})_{c}=0$. Introducing $\tau$ and defining $\omega\equiv \frac{v}{v_{c}}-1$, the expression (\ref{expansion}) reads \citep{Majhi:2016txt}
\begin{equation}
P=P_c+R \epsilon+B \epsilon \omega+D \omega^3+K \epsilon^2,
\end{equation}
where $R$, $B$, $D$, and $K$ are constants calculated from the derivatives at the critical point. It should be noted that higher-order terms have been discarded since they are very small. So, expanding (\ref{eosip}) around $\{\omega,\epsilon\}=\{0,0\}$ one obtains
\begin{equation}\label{expansion1}
    {P}=P_{c}+4\epsilon -6\omega\epsilon-\omega^{3}+9\omega^{2}\epsilon+\mathcal{O}(\omega^{4}).
\end{equation}
As ${P}$ is linear in ${T}$, then there is no higher order in $\epsilon$. Now, the procedure dictates using Maxwell's construction along with (\ref{expansion1}) to determine the small ${v}_{s}$ and large ${v}_{l}$ volumes in analytical way. Nevertheless, as this is not a genuine first order phase transition, and given that the unstable states are not confined, we cannot apply Maxwell's construction to determine the small and large volumes at the coexistent locus. Therefore, it is not possible to estimate the behavior of the molar volume in terms of the reduced temperature $\epsilon$, namely, the critical exponent $\hat{\beta}$ cannot be determined for this cosmological model. Next, from the Eq. (\ref{expansion1}), retaining terms up to first one obtains
\begin{equation}
\left(\frac{\partial {P}}{\partial {v}}\right)\bigg{|}_{\epsilon} \simeq \frac{B}{{v}_c} \epsilon,
\end{equation}
where the fact $\partial\omega/\partial{v}=1/v_{c}$ has been used. Then, the third expression in (\ref{exponent}) becomes 
\begin{equation}
\kappa_T \simeq \frac{1}{B \epsilon} \sim \epsilon^{-1},
\end{equation}
leading to $\hat{\gamma}=1$. Finally, evaluating (\ref{expansion1}) at ${T}=T_{c}$ (or equivalently $\epsilon=0$) one has
\begin{equation}
{P}-P_{c} \sim \omega^3 \sim\left(\frac{v}{v_{c}}-1\right)^3,
\end{equation}
providing $\hat{\delta}=3$. In summary, the following critical exponent has been obtained for the case $n=5$: $\{\hat{\alpha},\hat{\gamma},\hat{\delta}\}=\{0,1,3\}$. These coincide with those obtained for a Van der Waals gas under the mean field theory approach. It seems these values are in some sense ``universal". Nevertheless, it has been proven that certain BH systems have different values for their critical exponent \citep{Dehghani:2022gwg} (and references therein), what is more, here it is not possible to determine $\hat{\beta}$. Therefore, it is not possible to determine the critical exponent for the normalized curvature too.

\section{Topological interpretation}\label{sec5}

To further understand the thermodynamics behavior of the FLRW Universe in the EGB scenario, we are going to explore its topological interpretation following Duan's procedure \cite{Duan:1984ws}. Here, the main concept related to defects is the topological charge. In order to analyze the thermodynamic topology in this case, we need to modify some of the current definitions given in this analysis \cite{Wei:2021vdx,Wei:2022dzw}. This is a necessary step if we want to correctly compute the topological charge and use it to identify the topological classes.  
To do so, we need to construct the $\phi$ mapping as follows \cite{Wei:2021vdx,Wei:2022dzw}
\begin{equation}\label{phi}
\phi=\left(\phi^r, \phi^{\Theta}\right)=\left(\frac{\partial \mathcal{F}}{\partial r_{+}},-\cot \Theta \csc \Theta\right),
\end{equation}
where $\mathcal{F}$ is the off-shell free energy, given originally by
\begin{equation}
    \mathcal{F}=M-\frac{S}{\tau}.
\end{equation}
Nevertheless, in our case the mass (energy), should be replaced by the MS energy (\ref{MSEnergy}) evaluated at the $AH$, and the entropy $S$ corresponds in this situation to (\ref{entropy2}). It is worth mentioning that compared to the BH scenario, where the energy corresponds to the gravitational mass $M$ (expressed in terms of the event horizon $r_{+}$), here we have the MS energy of the whole Universe at the $AH$. This is an important fact, because this model is not static/stationary. In addition, $\tau\rightarrow -\tau$ is consistent with the on-shell case where $\tau=1/T$. This is so because in this system the temperature is negative in nature. Therefore, the redefined or adapted free energy becomes
\begin{equation}
    \mathcal{F}=E+\frac{S}{\tau}.
\end{equation}

Here, in the $\phi^{\Theta}$  component, the trigonometric function is chosen so that one zero point of the vector field can always be
found at $\Theta=\frac{\pi}{2}$. The other zero point can also be found by simply solving the equation $\phi^{r}=0$, which always results
in $\tau=\frac{1}{T}$ when the system is on-shell. The basic topological property associated with the zero point or topological defect of a field is its winding
number or topological charge. In this work, we use Duan’s $\phi$ mapping technique \cite{Duan:1984ws} to calculate the winding
number. To find the topological charge we first determine the unit vector $n$ of the field in Eq. (\ref{phi}), which are
\begin{equation}\label{unit}
\begin{aligned}
n^1 & =\frac{\phi^r}{\sqrt{\left(\phi^r\right)^2+\left(\phi^{\Theta}\right)^2}}, \\
n^2 & =\frac{\phi^{\Theta}}{\sqrt{\left(\phi^r\right)^2+\left(\phi^{\Theta}\right)^2}}.
\end{aligned}
\end{equation}
For the vector field, a topological current can be constructed in the coordinate space $x^{\mu}=(\tau,r_{+},\Theta)$ as follows \cite{Duan:1984ws}
\begin{equation}\label{current}
j^\mu=\frac{1}{2 \pi} \epsilon^{\mu \nu \rho} \epsilon_{a b} \partial_\nu n^a \partial_\rho n^b.
\end{equation}
The fundamental conditions that have to be fulfilled by the normalized vector $n^{a}$ are
\begin{equation}
n^a n_a=1 \quad \text { and } \quad n^a \partial_\nu n^a=0 .
\end{equation}
The current given in Eq. (\ref{current}), is a conserved quantity, which can be verified by applying the current conservation
law
\begin{equation}
\partial_\mu j^\mu=0 .
\end{equation}
where we use the following properties of Jacobi tensor
\begin{equation}
\epsilon^{a b} J^\mu\left(\frac{\phi}{x}\right)=\epsilon^{\mu \nu \rho} \partial_\nu \phi^a \partial_\rho \phi^b.
\end{equation}
the topological charge $W$ is related to the $0^{\text{th}}$ component of the current density of the topological current
through the following relation
\begin{equation}
W=\int_{\Sigma} j^0 d^2 x=\sum_{i=1}^N \beta_i \eta_i=\sum_{i=1}^N w_i,
\end{equation}
where $w_i$ is the winding number around the zero point. Also, $\beta_{i}$ y $\eta_{i}$ are the Hopf index and the Brouwer degree,
respectively. The detailed derivation of the above formula can be referred to \cite{Wei:2021vdx,Wei:2022dzw}.
Hence, the topological charge is also nonzero only at the zero points of the vector field. To find the exact zero point
where the topological charge is to be calculated, we plot the unit vector field $n$ and find out the zero point at which
it diverges. The zero point always turns out to be $(\frac{1}{T},\frac{\pi}{2})$. 

\begin{figure}[H]
    \centering
           \includegraphics[width=0.35\textwidth]{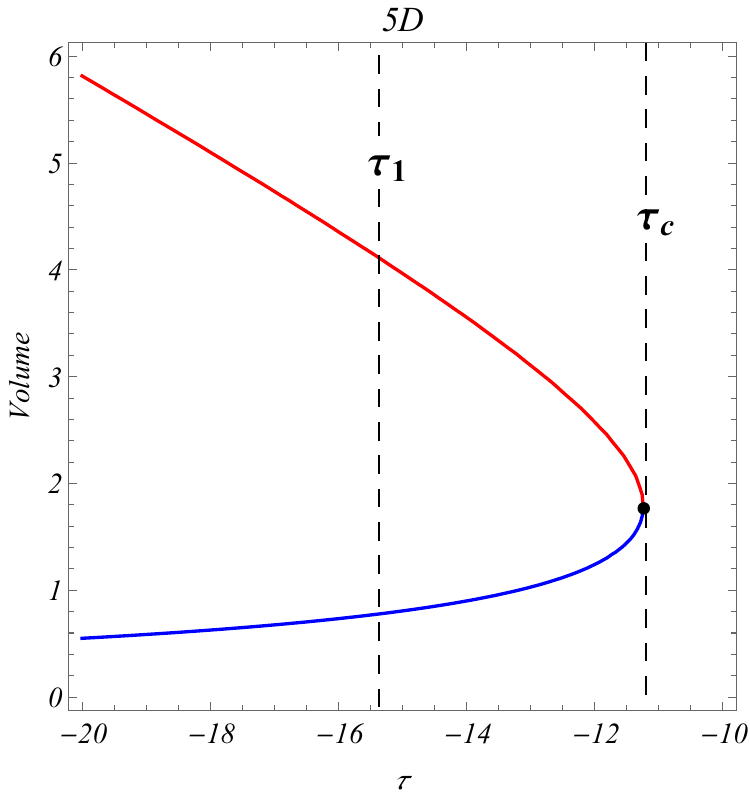} \  \includegraphics[width=0.35\textwidth]{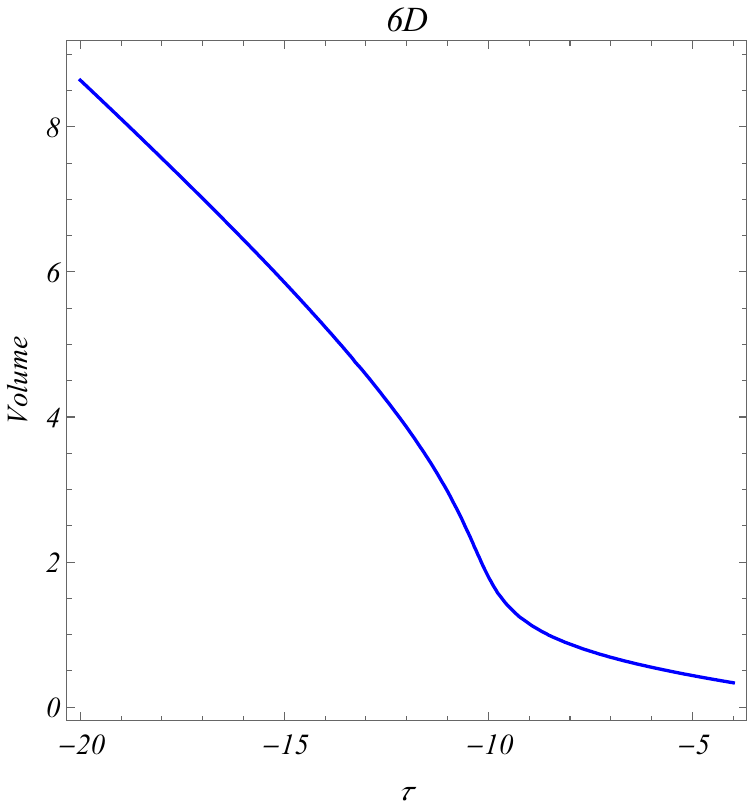}
    \caption{The $v-\tau$ plane for $n=5$ (left panel) and $n=6$ (right panel). For $n=5$ there is generating point, then phase transitions are happening, while for $n=6$ no generating point is present. For these plots we considered $\alpha=0.2$ square length. }
    \label{fig7}
\end{figure}

Now that the full picture is clear, we proceed in analyzing the topological thermodynamic behavior of this cosmological model. In this case the off-shell free energy $\mathcal{F}$ is given by
\begin{equation}
\mathcal{F}=\frac{ 2 \pi v\left[v^2+8(-3+n)(-2+n) \alpha\right]+\left[n-2\right]\left[v^2+4(n-4)(n-3) \alpha\right] \tau}{2^{n}\,\tau \,\Gamma\left[\frac{1}{2}(n-1)\right]\pi^{\frac{3n}{2}}v^{5n}},
\end{equation}
notice that we have replaced $\tilde{\alpha}$. From this expression and using the condition $\phi^{r}=0$ one gets
\begin{equation}
    \tau=
-\frac{\pi\left(v^2+16 \alpha\right)}{v},
\end{equation}
and 
\begin{equation}
\tau=-\frac{2 \pi\left(v^3+48 v \alpha\right)}{3\left(v^2+8 \alpha\right)},
\end{equation}
for $n=5$ and $n=6$, respectively. To obtain the critical values for $\tau$, that is, $\tau_{c}$, one needs to solve first from the above expressions $\partial \tau/\partial v=0$ for $v$, and then replace the value in the above expressions. For the corresponding dimensions one obtains
\begin{equation}
    v=4\sqrt{\alpha}, \quad (n=5), \quad \mbox{and} \quad v=
2 \sqrt{3 \alpha\pm i \sqrt{15} \alpha}, \quad (n=6).
\end{equation}
From this result we can notice two facts: i) the GB coupling $\alpha$ must be always positive if one wants to guarantee real solutions for $n=5$. Consequently phase transitions. ii) for $n=6$ there is not real solution. This fact corroborates that in this dimension no phase transition is occurring. To further realize this situation, we plotted in Fig. \ref{fig7} the molar volume versus $\tau$. The left panel shows two branches and one generating point (critical $\tau_{c}$). For $\tau_{1}=\tau<\tau_{c}$ we have a phase transition where the lower branch (blue line) corresponds to stable states and the upper one (red line) to unstable states. For values $\tau>\tau_{c}$ there is not phase transitions as expected. The right panel depicts the case $n=6$, clearly there is not a generating point, namely a critical values for $\tau$. Then, there is not phase transition as discussed early.

\begin{figure}[H]
    \centering
    \includegraphics[width=0.32\textwidth]{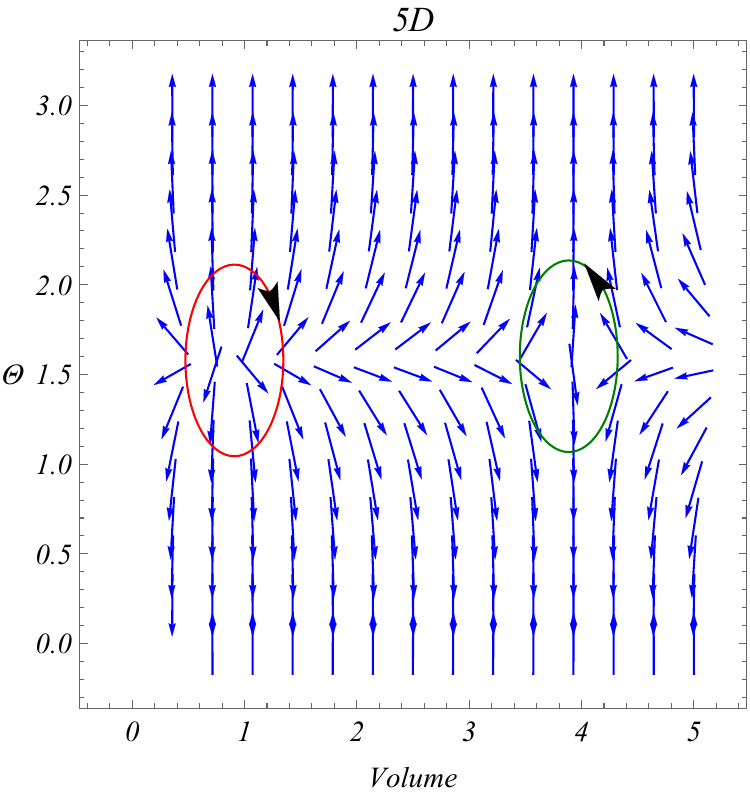} \  \includegraphics[width=0.32\textwidth]{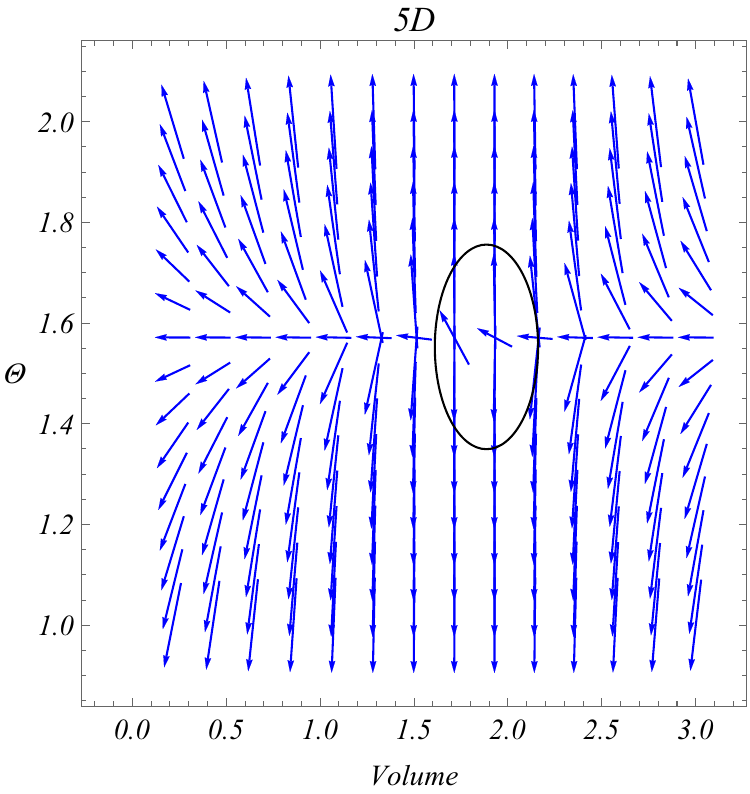}\
  \includegraphics[width=0.32\textwidth]{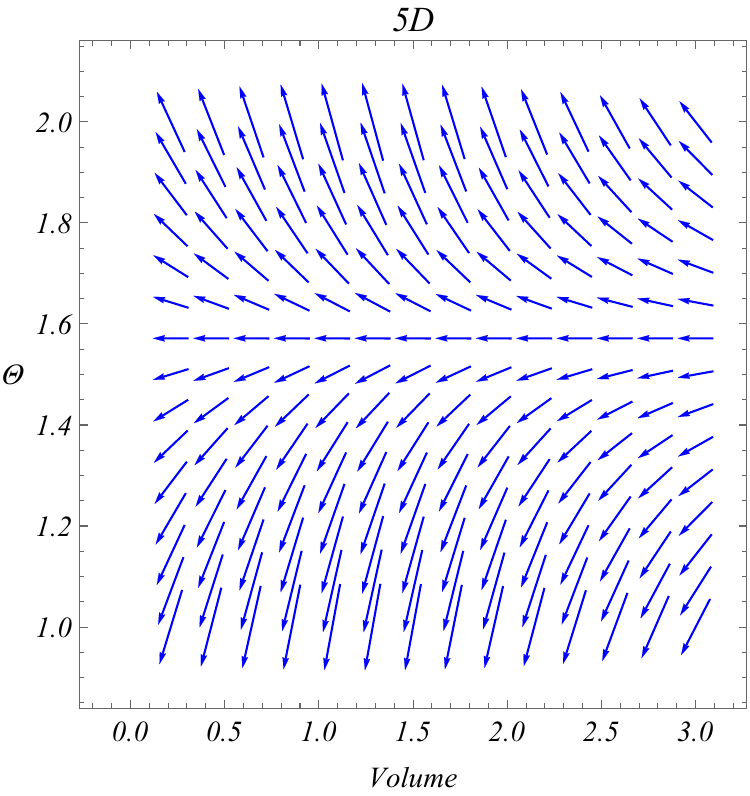}
    \caption{\textbf{Left panel}: The case $\tau<\tau_{c}$. The system has a total global topological charge equal to zero. \textbf{Central panel}: The case $\tau=\tau_{c}$. \textbf{Right panel}: The case $\tau>\tau_{c}$, no phase transition is present. For these plots we used $\alpha=0.2$ square length.}
    \label{fig6}
\end{figure}

\begin{figure}[H]
    \centering
           \includegraphics[width=0.35\textwidth]{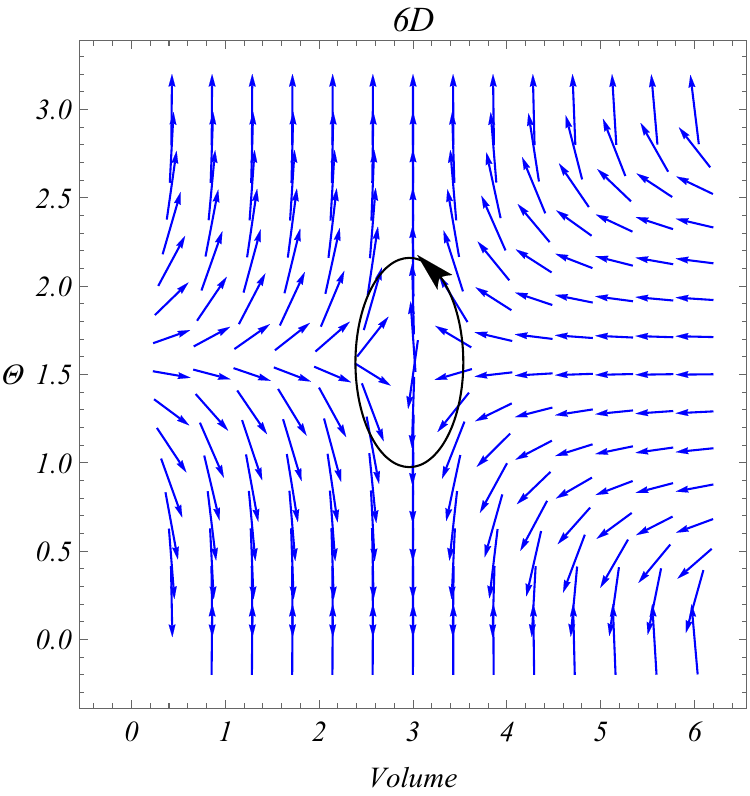}
    \caption{ The $\Theta-v$ plane for $n=6$. The system does not present any topological defect. In this case $W=+1$. For this plot we used  $\alpha=0.2$ square length.}
    \label{fig7-1}
\end{figure}

Fig. \ref{fig6} shows the $\Theta-v$ plane for the $n=5$ case. The left panel exhibits the situation for $\tau<\tau_{c}$. Here, as before we need to invert the interpretation. So, the loop enclosing arrows moving away, correspond to unstable states, with a topological charge (winding number) -1, while the green loop enclosing arrows approaching describe stable states with winding number +1. So, this system has a global topological charge $W=0$. The central panel, shows the critical case (generating point in the left panel of Fig. \ref{fig7}) and the right panel displays the case $\tau>\tau_{c}$ where no phase transition is occurring. The situation is quite different for the case $n=6$ depicted in Fig. \ref{fig7-1}, where no phase transition is happening. In this case the system has a global topological charge $W=+1$.

\section{Conclusions}\label{sec6}

This research has performed a thermodynamics, geometrothermodynamics and topological study for an FLRW Universe in the realm of Einstein-Gauss-Bonnet gravity theory. In this regard, the following main aspects and results of the present investigation can be highlighted.
\begin{itemize}
\item After a careful analysis about the matter content driving the FLRW model in the Einstein-Gauss-Bonnet scenario, we conclude that this hypothetical higher-dimensional universe is filled with a stiff matter distribution. This point is quite relevant in this study because in this way it is determined that this toy model is being described by an outer-past apparent horizon in an expanding era. This helps to determine the final sign of the temperature, negative in nature in this case.
    \item The system naturally constraint the space-time dimensions to be $n=5$ (lower bound) and $n=6$ (upper bound), for possible occurrence of phase transitions. 
    \item The case $n=5$ presents an ``inverted'' first order phase transition, where the swallow-tail exhibited by the Gibbs free energy is not physical. This is so because it is not possible to minimize it. For $n=6$, the system is not presenting phase transitions, in that case there are well-defined unstable/stable branches. The difference between these dimensions could lie on the fact that for $n=5$ the Gauss-Bonnet term is completely dynamical and for $n=6$ there is also a topological contribution. Then, preventing the system to undergo phase transitions.
    \item The geometrodynamics scheme, showed that the system has attractive interactions for $R_{N}>0$ and repulsive ones for $R_{N}<0$. The interpretation here has been inverted, since our model is not undergoing a typical first order phase transition. For $n=5$ there is a change in sign for the normalized curvature accounting for this unusual phase transitions, while for $n=6$ this change in sign only shows the stable/unstable branches. 
    \item The normalized curvature exhibits the usual divergence behavior at the critical point. However, one cannot interpret this divergence occurring at the spinodal curve. Despite throughout the manuscript we called it spinodal, for this model this curve is not delimiting the small and large volumes where the phase transition takes place. 
    \item As this is unusual first order phase transition, where unstable states dominates the model, the Maxwell's construction is not possible. Therefore, we cannot determine the critical exponent $\hat{\beta}$. However, we determined $\{\hat{\alpha},\hat{\gamma},\hat{\delta}\}=\{0,1,3\}$, as the mean field theory predicts. 
    \item The topological approach further corroborates the two previous analysis. In this concern, we find a generating point for $n=5$ that accounts for the phase transition. In this case, the system has a global topological charge equal to zero while for $n=6$, the scheme does not show any generating point; therefore, no phase transitions are occurring. In this case, it provides a global topological charge equal to one.
\end{itemize}

Despite the fact that the model does not show a realistic scenario, we can conclude that gravity theories formulated in dimensions higher than four, including high-order gravitational operators, offer a good scenario to test their feasibility (at least from the theoretical point of view) in the cosmological context by analyzing their thermodynamics description, microstructure, and topological behavior. Of course, as in the present case, they can reveal intriguing and interesting phenomena. It is suggested to explore the same study reported here for higher-order Lovelock terms, pure Lovelock gravity, or unified gravity theories such as Kaluza-Klein, for instance. Additionally, including a scalar field action for the action is an interesting issue for EGB gravity. The scalar field profoundly impacts the theory under consideration; for example, in four dimensions, it is well known that the EGB term is a topological invariant term and does not contribute to cosmological dynamics through the equation of motion. {Including the scalar field dramatically changes this situation; in this case, the EGB term does contribute to the dynamics with a rich non-linear structure, and therefore, in an FLRW universe, it could be possible to find phase transition as we describe in the present article. Also, we need to study what kind of scalar function we are considering in higher dimensions to obtain the cosmological behavior and, therefore, the subsequent thermal history of the universe where, of course, we hope to modify our results on phase transition. These modifications are mainly supported because there is a strong suggestion about the relationship between the first law of thermodynamics of the apparent horizon and the Friedmann equation. To our knowledge, a critical point is that no one has studied the apparent thermodynamics for the case where the geometry is coupled with a single scalar field. We left the interesting model with the EGB term and the inclusion of scalar field coupling as a future work.  Also, we did not address the issue of examining the tensor perturbation EGB theories in light of the GW170817 results, as discussed in the references \cite{Oikonomou:2024etl}, \cite{Oikonomou:2021kql}, \cite{Oikonomou:2025ccs} and \cite{Odintsov:2025kyw} . Those results and their effect on the phase transition pattern could be essential for testing our model and its theoretical predictions, and we will discuss those points elsewhere.}

\section*{ACKNOWLEDGEMENTS}
J. Saavedra and F. Tello-Ortiz acknowledge to grant FONDECYT N°1220065, Chile.
F. Tello-Ortiz acknowledges VRIEA-PUCV
for financial support through Proyecto Postdoctorado 2023 VRIEA-PUCV. The authors are grateful for fruitful discussions with Ramón Herrera, Miguel Cruz, and Manuel González during the manuscript preparation.

\bibliography{biblio.bib}
\bibliographystyle{elsarticle-num}

%\printbibliography

\end{document}